\documentclass[11pt,a4paper]{article}
\pdfoutput=1

\usepackage[intlimits]{amsmath}
\usepackage{amsfonts}
\usepackage{amssymb}
\usepackage{mathrsfs}
\usepackage{amsthm}
\usepackage[english]{babel}
\usepackage[utf8]{inputenc}
\usepackage[T1]{fontenc}
\usepackage{epsfig}
\usepackage{textcomp}
\usepackage{authblk}
\usepackage{simplewick}

\def\a{\alpha}

\def\G{\Gamma}
\def\d{\delta}

\def\z{\zeta}

\def\th{\theta}

\def\La{\Lambda}

\def\s{\sigma}

\def\t{\tau}
\def\f{\phi}

\def\vf{\varphi}

\def\ps{\psi}
\def\o{\omega}

\newcommand{\ti}[1]{\tilde{#1}}

\newcommand{\barg}{\bar{g}}
\newcommand{\barn}{\bar{n}}

\newcommand{\bars}{\bar{\sigma}}
\newcommand{\bart}{\bar{\tau}}
\newcommand{\barG}{\bar{\Gamma}}

\newcommand{\hH}{\hat{H}}
\newcommand{\hU}{\hat{U}}
\newcommand{\hS}{\hat{S}}

\newcommand{\hf}{\hat{\phi}}
\newcommand{\hps}{\hat{\psi}}
\newcommand{\hvf}{\hat{\varphi}}
\newcommand{\hpi}{\hat{\pi}}
\newcommand{\hr}{\hat{\rho}}
\newcommand{\ha}{\hat{a}}
\newcommand{\hq}{\hat{q}}
\newcommand{\hp}{\hat{p}}
\newcommand{\hQ}{\hat{Q}}
\newcommand{\hP}{\hat{P}}

\newcommand{\hV}{\hat{V}}

\providecommand{\abs}[1]{\lvert#1\rvert}
\providecommand{\babs}[1]{\big\lvert#1\big\rvert}
\providecommand{\Babs}[1]{\Big\lvert#1\Big\rvert}

\providecommand{\norm}[1]{\lVert#1\rVert}
\providecommand{\bnorm}[1]{\big\lVert#1\big\rVert}

\newcommand{\de}{\partial}
\newcommand{\half}{\frac{1}{2}}

\newcommand{\p}{\prime}
\newcommand{\beq}{\begin{equation}}
\newcommand{\eeq}{\end{equation}}
\newcommand{\beqnn}{\begin{equation*}}
\newcommand{\eeqnn}{\end{equation*}}
\newcommand{\bea}{\begin{eqnarray}}
\newcommand{\eea}{\end{eqnarray}}
\newcommand{\nn}{\nonumber}

\newcommand{\sgn}{\mathrm{sgn}}
\newcommand{\vecx}{\vec{x}}
\newcommand{\vecy}{\vec{y}}
\newcommand{\vecz}{\vec{z}}
\newcommand{\veck}{\vec{k}}

\newcommand{\dg}{\dagger}

\newcommand{\scn}{\textsc{n}}

\newcommand{\mcal}[1]{\mathcal{#1}}

\newcommand{\mcalR}{\mathcal{R}}

\newcommand{\mcalU}{\mathcal{U}}
\newcommand{\hmcalQ}{\hat{\mathcal{Q}}}
\newcommand{\hmcalP}{\hat{\mathcal{P}}}
\newcommand{\mcalH}{\mathcal{H}}
\newcommand{\hmcalH}{\hat{\mathcal{H}}}

\newcommand{\mscr}[1]{\mathscr{#1}}

\newcommand{\mscrR}{\mathscr{R}}

\newcommand{\mscrS}{\mathscr{S}}
\newcommand{\mscrH}{\mathscr{H}}
\newcommand{\mbbN}{\mathbb{N}}
\newcommand{\mbbZ}{\mathbb{Z}}

\newcommand{\mbbR}{\mathbb{R}}
\newcommand{\mbbC}{\mathbb{C}}

\numberwithin{equation}{section}

\begin{document}

\title{On L\'{e}on van Hove's 1952 article on the foundations of Quantum Field Theory \\[4mm]
\large O artigo de L\'{e}on van Hove de 1952 sobre os fondamentos da Teoria Qu\^{a}ntica de Campos}

\author{\Large{Fulvio Sbis\`{a}\footnote{fulviosbisa@gmail.com}}}

\affil{\normalsize{Departamento de F\'{i}sica Te\'{o}rica, Universidade do Estado do Rio de Janeiro,\\
20550-013, Rio de Janeiro, RJ, Brazil}}

\date{}

\vspace{.5cm}

\maketitle

\thispagestyle{empty}

\begin{abstract}
In 1952, L\'{e}on van Hove published an article, in French, with the title \emph{Les dif\mbox{}f\mbox{}icult\'{e}s de divergences pour um mod\`{e}le particulier de champ quantif\mbox{}i\'{e}}. The article is frequently cited in relation to Haag's theorem and to the issue of the existence of unitarily inequivalent representations of the canonical commutation relations in Quantum Field Theory. Summarizing in brief, it suggests a link between the appearance of divergences in perturbative Quantum Field Theory and the fact that quantum states belonging to an interacting theory do not belong to the same Hilbert space of the free theory. It also suggests that renormalization fails to provide an accurate description of the time evolution of the quantum f\mbox{}ield, although it correctly accounts for the $S$ matrix. Due to its relevance, and to the dif\mbox{}f\mbox{}iculty of f\mbox{}inding an English translation, the ideas contained in this article are proposed again here, expanded with derivations and accompanied by a discussion aimed at putting the analysis into context. We highlight the main points from the perspective of a contemporary reader, and underline the dif\mbox{}ferences with the standard approach usually taught in curricular courses in Quantum Field Theory.\\
\textbf{Keywords:} Quantum Field Theory, Perturbative divergences, Renormalization.

\bigskip

Em 1952, L\'{e}on van Hove publicou um artigo, em Franc\^{e}s, com o t\'{i}tulo \emph{Les dif\mbox{}f\mbox{}icult\'{e}s de divergences pour um mod\`{e}le particulier de champ quantif\mbox{}i\'{e}}. O artigo \'{e} frequentemente citado em rela\c{c}\~{a}o ao teorema de Haag e \`{a} quest\~{a}o da exist\^{e}ncia de representa\c{c}\~{o}es unit\'{a}riamente n\~{a}o equivalentes das rela\c{c}\~{o}es de comuta\c{c}\~{a}o can\^{o}nica na Teoria Qu\^{a}ntica de Campos. Resumindo, ele sugere uma liga\c{c}\~{a}o entre o aparecimento de diverg\^{e}ncias na Teoria Qu\^{a}ntica de Campos perturbativa e o fato de que estados qu\^{a}nticos pertencentes a uma teoria interagente n\~{a}o pertencem ao mesmo espa\c{c}o de Hilbert da teoria livre. Tamb\'{e}m sugere que a renormaliza\c{c}\~{a}o falha ao fornecer uma descri\c{c}\~{a}o precisa da evolu\c{c}\~{a}o temporal do campo qu\^{a}ntico, apesar de explicar corretamente a matriz $S$. Devido \`{a} sua relev\^{a}ncia e \`{a} dif\mbox{}iculdade de encontrar uma tradu\c{c}\~{a}o, as id\'{e}ias contidas neste artigo s\~{a}o propostas novamente aqui, ampliadas com deriva\c{c}\~{o}es e acompanhadas de uma discuss\~{a}o que visa contextualizar a an\'{a}lise. Destacamos os pontos principais da perspectiva de um leitor contempor\^{a}neo e destacamos as diferen\c{c}as com a abordagem padr\~{a}o normalmente ensinada nos cursos curriculares da Teoria Qu\^{a}ntica de Campos.\\
\textbf{Palavras-chave:} Teoria Qu\^{a}ntica de Campo, diverg\^{e}ncias, renormaliza\c{c}\~{a}o.
\end{abstract}

Shortly after the birth and systematization of renormalization \cite{Dyson 1951}, the Belgian physicist L\'{e}on van Hove wrote two articles (\cite{van Hove 1951} in 1951, and \cite{van Hove 1952} in 1952) on the formal structure of Quantum Field Theory (QFT). The aim of the second article \cite{van Hove 1952} in particular was to shed light on the reason behind the appearance of the divergences which plague QFT, and to assess whether renormalized perturbative QFT can be considered an accurate description of a non-perturbative QFT. The abstract reads as follows:
\begin{quotation}
It is well known that for a neutral scalar f\mbox{}ield in scalar interaction with inf\mbox{}initely heavy, f\mbox{}ixed point sources, the stationary states can be determined exactly. This simple model of quantum f\mbox{}ield is considered for the discussion of the following two problems: to investigate the origin of the divergences which are unavoidably brought in when the interaction is treated as a perturbation; and to see how good a description of the exact solution is obtained from the perturbative approach, as improved by the renormalization technique to discard the divergences. The origin of the divergences is found to lie in the fact that the stationary states of the f\mbox{}ield interacting with the sources are not linear combinations of the stationary states of the free f\mbox{}ield. The former are not contained in the Hilbert space spanned by the latter (they even turn out to be orthogonal to this space). As a consequence of the results obtained in a previous article, a similar property is shown to hold for the more realistic case of two interacting f\mbox{}ields under the mere assumption that stationary states exist in presence of interaction. Regarding the second problem, whereas the exact solution and the results obtained by perturbation and renormalization methods are in agreement for the $S$ matrix, they are found to disagree for the unitary matrix $S(t)$ expressing the change of the wave vector between the times $t = - \infty$ and $t$ f\mbox{}inite.
\end{quotation}

Despite being more than sixty years old, this article is still cited nowadays, mainly in connection with Haag's theorem and the existence of unitarily inequivalent representations of the canonical commutation relations. Nevertheless, it is largely unknown to the physicists community, apart from those who work on the mathematical aspects of QFT. It is in our opinion important, both from a historic and a didactic point of view, to give the ideas contained in this article the visibility they deserve. Even more so since English translations of the French original are very hard to f\mbox{}ind, if they exist at all. We therefore propose again these ideas here, by no means aiming to provide a substitute to the original. 

Apart from a slight reorganization, we follow the lines of the article (sections \ref{sec: stationary states} to \ref{sec: S matrix and renormalization}) adding explanations and derivations to improve the understandability of the analysis. In the process we corrected one typo and a wrong claim, and switched to a notation closer to the contemporary taste. To help place the results into context, we added a ``Comments'' section (number \ref{sec: Comments}) where we also emphasize the points which we believe are more interesting to the contemporary reader, and highlight how and where the article departs from the standard exposition of QFT a student is typically exposed in curricular courses at the undergraduate level.

A notable dif\mbox{}ference with the original, which considers the general case of $\scn \geq 1$ point-like sources, is that we focused on the case $\scn = 2$ where only two sources are present. We believe that this way the results of the article stand out more clearly, more than compensating the loss in generality. Nevertheless, we comment where appropriate on how the particular case relates to the general one. We remark that, apart from the abstract which appeared also in English in the original paper, when we report van Hove's words we are actually reporting our own translation of them.

\section{Introduction}
\label{sec: introduction}

The contemporary outlook on renormalization has been greatly inf\mbox{}luenced, besides its celebrated empirical success, by the role it had in the development of the Standard Model \cite{Weinberg Nobel} and by the conceptual re-elaboration operated by Kenneth Wilson \cite{Wilson Kogut 1974, Wilson 1975, Wilson Nobel}. Indeed, the application to Quantum Field Theory of the ideas of the renormalization group and the contextualization of QFT into the framework of ef\mbox{}fective f\mbox{}ield theory has elevated renormalization to the status of a physical procedure, when previously it was at times regarded as a mere technique to ``get rid of divergences by hiding them under the carpet''.

However, in the early days of QFT the feel about renormalization was much more varied, ranging from the enthusiasm of the pioneers of renormalization to the severe criticism of others (including physicists as notable as Dirac, Heisenberg and Landau) \cite{Cao Schweber Synthese 1993}. In the introduction to \cite{van Hove 1952} (to which hereafter we refer simply as ``the article''), van Hove expresses his point of view on renormalization (at any rate, his point of view at the time) saying that ``this method, which absorbs the divergences into the fundamental constants of the theory, cannot be regarded as def\mbox{}initive, and the search for a more fundamental method, as well as the comparison of the latter with the method of perturbative renormalization, seem indispensable tasks''. Well aware of the dif\mbox{}f\mbox{}iculties involved in this search, he suggests as an interesting direction to be pursued the study of simplif\mbox{}ied models which are simple enough to admit a rigorous (or even exact) treatment, but nevertheless share in a weaker form the divergences characteristic of the general cases. This point of view had already been advocated by Sollfrey and Goertzel \cite{Sollfrey 1951}, who studied the mechanical model where a vibrating string is coupled to a harmonic oscillator.

The simplif\mbox{}ied model chosen by van Hove for his analysis was instead that of a neutral scalar f\mbox{}ield coupled to point-like sources which are too massive to move as a result of the interaction. Notably, if the positions of the sources are held strictly f\mbox{}ixed then the equation of motion for the scalar f\mbox{}ield can be solved exactly. The present contribution is devoted to revive van Hove's analysis of this model, taking advantage of the benef\mbox{}its of hindsight. The exposition is structured as follows: after an introduction on the canonical commutation relations and the interaction picture in section \ref{sec: CCR interaction picture}, the exact solution of van Hove's model is derived in section \ref{sec: stationary states}, and a general theorem about the structure of the space of state vectors (and a generalization thereof) is proved in section \ref{sec: space of the state vectors}. These results are used in section \ref{sec: QFT and perturbative} to try to understand why the perturbative approach inevitably leads to the problem of divergences. Finally, applying the procedure of perturbative renormalization to the model and comparing the result with the exact one, in section \ref{sec: S matrix and renormalization} it is discussed until which point the perturbative renormalization provides a satisfactory approximation of the exact solution. Some personal comments and criticisms are proposed in section \ref{sec: Comments}, and the conclusions are summarized in section \ref{sec: Conclusions}.

\section{Commutation relations and interaction picture}
\label{sec: CCR interaction picture}

The notion of ``interaction picture'' is a familiar one from the curricular courses in non-relativistic Quantum Mechanics (QM) and QFT, and lies at the basis of the perturbative formulation of the latter. Nevertheless, the subtleties involved in its introduction for systems with an inf\mbox{}inite number of degrees of freedom are not always (or not at all) adequately discussed, and are of central importance in the article. Furthermore, there is an interplay between this notion and the choice of representation of the canonical commutation relations, which it is helpful to spell out. Before starting with the exposition of the article, we therefore spend some words to clarify these points.

\subsection{The interaction picture}

The general idea underlying the interaction picture is that, in many situations, instead of concentrating directly on the solution of a model it is more convenient to investigate its behavior relatively to a set of known solutions. This may happen because the equations are easier to solve from the ``relative'' point of view, or because the physics of the model becomes more transparent.

\subsubsection{Dirac's interaction picture}

In the context of quantum physics, the interaction picture was introduced by Dirac \cite{Dirac 1926} to study the absorption and stimulated emission of radiation by an atom (with spontaneous emission treated in \cite{Dirac 1927}). The procedure closely parallels the method of ``variation of parameters (or constants)'', well-known from the theory of linear dif\mbox{}ferential equations \cite{Arnold 1992}.\footnote{In classical mechanics, this method proved very useful to study the inf\mbox{}luence of other planets on the motion of one chosen planet around the sun, providing an example of the usefulness of the ``relative'' approach.} In this section, as usual we follow von Neumann \cite{von Neumann 1932} and assume that the states of a quantum system are described by elements of a complex, separable Hilbert space $\mscrH$, and that the observables are described by self-adjoint (linear) operators on $\mscrH$.\footnote{See \cite{Helmberg 1969} for a very nice introduction to Hilbert spaces and the spectral theorem.} We use Dirac's notation $\langle \, \mid \, \rangle$ to indicate the inner product in $\mscrH$, and indicate with $[ \, , \, ]$ the commutator.

Consider then a time-independent Hamiltonian $\hH_{_{0}}$ (which may for example describe an isolated atom or molecule), and let $(\ps_{n})_{n}$ be the set of its (orthonormalized) eigenstates and $(E_{n})_{n}$ its eigenvalues, supposing for simplicity the spectrum to be purely discrete and non-degenerate. The generic normalized solution of the Schr\"{o}dinger equation
\begin{equation}
i \hbar \, \de_{t} \, \ps = \hH_{_{0}} \, \ps
\end{equation}
is then
\begin{equation}
\ps(t) = \sum_{n = 0}^{\infty} c_{n} \, e^{-\frac{i}{\hbar} (t - t_{_{0}}) E_{n}} \, \ps_{n} \quad ,
\end{equation}
with $\sum_{n} \, \abs{c_{n}}^{2} = 1\,$. The reference time $t_{_{0}}$ can be chosen freely, and is the time at which $c_{n}$ coincides with the projection of $\ps$ on $\ps_{n}\,$, so it follows that
\begin{equation}
\ps(t) = e^{-\frac{i}{\hbar} (t - t_{_{0}}) \hH_{_{0}}} \, \ps(t_{_{0}}) \quad .
\end{equation}

Suppose now this system is perturbed, for example by an external electromagnetic f\mbox{}ield, and call $\hH_{_{\textup{int}}}$ the (possibly time-dependent) Hamiltonian describing the interaction, so the total Hamiltonian is $\hH = \hH_{_{0}} + \hH_{_{\textup{int}}}$. Since the collection of states $(\ps_{n})_{n}$ generates the Hilbert space of state vectors of the system, the generic solution of the interacting Schr\"{o}dinger equation
\begin{equation} \label{Schrodinger picture states eq}
i \hbar \, \de_{t} \, \ps = \big( \hH_{_{0}} + \hH_{_{\textup{int}}}(t) \big) \, \ps
\end{equation}
can be expanded on this basis, and therefore can be written in the form
\begin{equation} \label{perturbed Schrodinger sol}
\ps(t) = \sum_{n = 0}^{\infty} c_{n}(t) \, e^{-\frac{i}{\hbar} (t - t_{_{0}}) E_{n}} \, \ps_{n} \quad .
\end{equation}
It is apparent that the constants (or parameters) $(c_{n})_{n}$ of the unperturbed solution are turned into the functions of time $(c_{n}(t))_{n}$, which describe the exact interacting solution relatively to the free one. The description in terms of the $(c_{n}(t))_{n}$ is particularly convenient if the interaction is non-vanishing only in a f\mbox{}inite time interval $[ \, t_{\textup{i}} \, , t_{\textup{f}} \, ]$. Supposing the system is in the energy level $E_{j}$ before the interaction kicks in, then the coef\mbox{}f\mbox{}icient $c_{n}(t_{\textup{f}})$ gives the transition amplitude from the energy level $E_{j}$ to the energy level $E_{n}$ caused by the interaction, so the ``relative'' description in terms of the $(c_{n}(t))_{n}$ is very transparent from a physical point of view.

\subsubsection{The pictures in Quantum Mechanics}

Although very convenient to study transition amplitudes, the description in terms of the functions $(c_{n}(t))_{n}$ is less suited for other types of manipulations. One would like then to describe the system with an abstract state function whose evolution is nevertheless determined only by the Hamiltonian which describes the interaction, retaining the relative character of the description. It is indeed not dif\mbox{}f\mbox{}icult to verify that the state function
\begin{equation} \label{int picture wavefunction def}
\ps_{_{\textup{I}}}(t) = \sum_{n = 0}^{\infty} c_{n}(t) \, \ps_{n} \quad ,
\end{equation}
obtained by ``undoing'' the time evolution of the free energy eigenstates, satisf\mbox{}ies the equation
\begin{equation} \label{int picture states eq}
i \hbar \, \de_{t} \, \ps_{_{\textup{I}}} = \hH_{_{\textup{I}}}(t) \, \ps_{_{\textup{I}}} \quad ,
\end{equation}
where
\begin{equation}
\hH_{_{\textup{I}}}(t) = e^{+ \frac{i}{\hbar} (t - t_{_{0}}) \hH_{_{0}}} \,\, \hH_{_{\textup{int}}}(t) \,\, e^{- \frac{i}{\hbar} (t - t_{_{0}}) \hH_{_{0}}} \quad .
\end{equation}
Note that, indicating with $\ps_{_{\textup{S}}}$ the state function which was denoted with $\ps$ in the equations (\ref{Schrodinger picture states eq})--(\ref{perturbed Schrodinger sol}), the relation (\ref{int picture wavefunction def}) can be written abstractly as
\begin{equation} \label{int picture states def}
\ps_{_{\textup{I}}}(t) = e^{+ \frac{i}{\hbar} (t - t_{_{0}}) \hH_{_{0}}} \, \ps_{_{\textup{S}}}(t) \quad .
\end{equation}
Therefore, $\ps_{_{\textup{I}}}$ is a good candidate for the ``relative'' state function we were looking for. To complete the quantum description, it is necessary to provide a way to compute the expectation values of the observables directly from $\ps_{_{\textup{I}}}\,$. Consider for example an observable $\mcal{O}$, described by the operator $\hat{\mcal{O}}_{_{\!\textup{S}}}$ in the formulation where the state obeys (\ref{Schrodinger picture states eq}). Since the (free evolution) operator $e^{- \frac{i}{\hbar} (t - t_{_{0}}) \hH_{_{0}}}$ is unitary,\footnote{A map $\mcalU$ is said to unitary if $\mcalU \mcalU^{\dag} = \mcalU^{\dag} \mcalU = \textup{Id}\,$, where $\textup{Id}$ denotes the identity operator in the appropriate Hilbert space.} with inverse $\big( e^{- \frac{i}{\hbar} (t - t_{_{0}}) \hH_{_{0}}} \big)^{-1} = \big( e^{- \frac{i}{\hbar} (t - t_{_{0}}) \hH_{_{0}}} \big)^{\dag} = e^{+ \frac{i}{\hbar} (t - t_{_{0}}) \hH_{_{0}}}$, this can be achieved by mapping $\hat{\mcal{O}}_{_{\!\textup{S}}}$ into a (time-dependent) operator $\hat{\mcal{O}}_{_{\!\textup{I}}}(t)$ by the relation
\begin{equation} \label{Interaction picture observable def}
\hat{\mcal{O}}_{_{\!\textup{I}}}(t) = e^{+ \frac{i}{\hbar} (t - t_{_{0}}) \hH_{_{0}}} \,\, \hat{\mcal{O}}_{_{\!\textup{S}}} \,\, e^{- \frac{i}{\hbar} (t - t_{_{0}}) \hH_{_{0}}} \quad .
\end{equation}
The expectation values are trivially left unaltered by the simultaneous transformation of state function and operator
\begin{equation}
\Big\langle \ps_{_{\textup{I}}}(t) \, \Big\vert \, \hat{\mcal{O}}_{_{\!\textup{I}}}(t) \, \Big\vert \, \ps_{_{\textup{I}}}(t) \Big\rangle = \Big\langle \ps_{_{\textup{S}}}(t) \, \Big\vert \, \hat{\mcal{O}}_{_{\!\textup{S}}} \, \Big\vert \, \ps_{_{\textup{S}}}(t) \Big\rangle \quad ,
\end{equation}
and the new operator obeys the evolution equation
\begin{equation} \label{int picture operator eq}
\frac{d}{dt} \hat{\mcal{O}}_{_{\!\textup{I}}}(t) = \frac{i}{\hbar} \, \Big[ \hH_{_{0}} \, , \hat{\mcal{O}}_{_{\!\textup{I}}}(t) \Big] \quad .
\end{equation}
Looking at (\ref{int picture states eq}) and (\ref{int picture operator eq}), it is apparent that in this relative description the dynamical problem is split into two parts: the time evolution of the state function is determined by the interaction part of the Hamiltonian, while the time evolution of the operators is determined by the free part of the Hamiltonian. 

The description of the dynamical problem in terms of $\ps_{_{\textup{S}}}(t)$ and $\hat{\mcal{O}}_{_{\!\textup{S}}}$, where the state function evolves according to (\ref{Schrodinger picture states eq}), is called the \emph{Schr\"{o}dinger picture}, while the description in terms of $\ps_{_{\textup{I}}}(t)$ and $\hat{\mcal{O}}_{_{\!\textup{I}}}(t)$, where the state function evolves according to (\ref{int picture states eq}) and the operators according to (\ref{int picture operator eq}), is called the \emph{interaction picture}. Another description, more used in QFT, is the so-called \emph{Heisenberg picture}, where the state function $\ps_{_{\textup{H}}}$ does not evolve in time and the dynamics is completely cast on the operators, whose time evolution is determined by the equation
\begin{equation} \label{Heisenberg picture operator eq}
\frac{d}{dt} \hat{\mcal{O}}_{_{\!\textup{H}}}(t) = \frac{i}{\hbar} \, \Big[ \hH(t) \, , \hat{\mcal{O}}_{_{\!\textup{H}}}(t) \Big] \quad .
\end{equation}
All these pictures provide equivalent descriptions of the system under study and of its time evolution. The (arbitrary) reference time $t_{_{0}}$ is the time when the three pictures coincide, the customary choice being $t_{_{0}} = 0\,$.

\subsection{The CCR and the unitary equivalence theorems}

It is apparent that the possibility of introducing the interaction picture relies completely on the assumption that the eigenstates of the free Hamiltonian $\hH_{_{0}}$ generate the space of state vectors of the interacting system. In other words, it relies on the assumption that the free Hamiltonian $\hH_{_{0}}$ can be exponentiated to a unitary operator on the space of states of the \emph{interacting} theory.

This is one of those assumptions that, in the early period of the new quantum theory,\footnote{We use the term ``new quantum theory'' to denote the development of QM starting around 1925 with Heisenberg's f\mbox{}irst paper \cite{Heisenberg 1925} on matrix mechanics.} were accepted as true even if fully f\mbox{}ledged mathematical proof were lacking. Other important examples are the equivalence, heuristically proved by Schr\"{o}dinger \cite{Schrodinger 1926}, between Heisenberg, Born and Jordan's matrix mechanics \cite{Heisenberg 1925, Born Jordan 1926, Born Heisenberg Jordan 1926} and Schr\"{o}dinger's wave mechanics \cite{Schrodinger collected}, and the assertion (explicit for example in Dirac \cite{Dirac 1926}) that the theory is completely determined by specifying at the abstract level the commutators between the observables.

\subsubsection{The Canonical Commutation Relations}

Indeed, considering a system of $n$ degrees of freedom where the canonical position and momentum observables are represented by the operators $\big( \hQ_{j}, \hP_{j} \big)_{j = 1 \, , \ldots , n}\,$, in both the Schr\"{o}dinger formulation and the Heisenberg-Born-Jordan formulation the following commutation relations hold
\begin{align} \label{CCR}
\Big[ \hQ_{j} \, , \hQ_{k} \Big] &= \Big[ \hP_{j} \, , \hP_{k} \Big] = 0 \quad , & \Big[ \hQ_{j} \, , \hQ_{k} \Big] &= i \hbar \, \d_{jk} \, \textup{Id} \quad ,
\end{align}
where $0$ and $\textup{Id}$ respectively denote the null and the identity operator. For the sake of simplicity, we limit ourselves here to the quantum mechanics of spin-zero particles, so all physically relevant quantities can be expressed in terms of the $\hQ$s and $\hP$s. The commutation relations (\ref{CCR}) are known as the Canonical Commutation Relations (CCR), or as the Heisenberg commutation relations (although it would historically more appropriate to name them after Dirac, Born and Jordan since they f\mbox{}irst appeared explicitly in \cite{Dirac 1925} and \cite{Born Jordan 1926} and not in \cite{Heisenberg 1925}).

As it is widely known, in Schr\"{o}dinger's approach the operators $\big( \hQ_{j}, \hP_{j} \big)_{\!j}$ are realized respectively as the multiplication and dif\mbox{}ferentiation operators
\begin{align}
\hQ_{j} &: \,\,\, \ps(x_{_{1}} \, , \ldots , x_{n}) \to x_{j} \, \ps(x_{_{1}} \, , \ldots , x_{n}) \quad , \label{Schrodinger Q} \\[1mm]
\hP_{j} &: \,\,\, \ps(x_{_{1}} \, , \ldots , x_{n}) \to i \hbar \, \frac{\de}{\de x_{j}} \, \ps(x_{_{1}} \, , \ldots , x_{n}) \quad ,  \label{Schrodinger P}
\end{align}
in the space $L^{2}(\mbbR^{n}, \mbbC)$ of complex-valued functions which are square-integrable on $\mbbR^{n}$, while in Heisenberg, Born and Jordan's approach they are realized as appropriate inf\mbox{}inite matrices acting on inf\mbox{}inite sequences of complex numbers. Among the founders, Dirac was probably the one who identif\mbox{}ied more clearly the algebraic structure of the new QM, which he formulated directly in terms of non-commuting abstract ``q-numbers'' obeying (\ref{CCR}) (the observables), while ordinary complex numbers were referred to as ``c-numbers''.

An important re-formulation of the kinematical structure of QM was given by Weyl \cite{Weyl 1928}, who proposed that the operators $\big( \hQ_{j}, \hP_{j} \big)_{\!j}$ should more appropriately be seen as the generators of the one-parameter unitary groups
\begin{align}
\hU^{(j)}(\s_{j}) &= e^{\frac{i}{\hbar} \s_{j} \hP_{j}} \quad , & \hV^{(j)}(\t_{j}) &= e^{\frac{i}{\hbar} \t_{j} \hQ_{j}} \quad ,
\end{align}
which, considering for simplicity the case $n = 1\,$, in the Schr\"{o}dinger representation act as follows
\begin{align}
\big( \hU(\s) \, \ps \big)(x) &= \ps (x + \s) & \big( \hV(\t) \, \ps \big)(x) &= e^{\frac{i}{\hbar} \t x} \, \ps(x) \quad .
\end{align}
In the general case, Weyl heuristically proved that the position and momentum operators generate two $n$-parameter unitary groups, $\hU \big( \boldsymbol{\s} \big)$ and $\hV(\boldsymbol{\t})$, which obey
\begin{align} \label{Weyl CCR 1}
\hU \big( \boldsymbol{\s} \big) \, \hU \big( \boldsymbol{\s}^{\p} \big) &= \hU \big( \boldsymbol{\s} + \boldsymbol{\s}^{\p} \big) \quad , & \hV(\boldsymbol{\t}) \, \hV(\boldsymbol{\t}^{\p}) &= \hV(\boldsymbol{\t} + \boldsymbol{\t}^{\p}) \quad ,
\end{align}
\begin{equation} \label{Weyl CCR 2}
\hU(\boldsymbol{\s}) \, \hV(\boldsymbol{\t}) = e^{\frac{i}{\hbar} \big( \s_{_{1}} \t_{_{1}} + \ldots + \s_{n} \t_{n} \big)} \, \hV(\boldsymbol{\t}) \, \hU(\boldsymbol{\s}) \quad ,
\end{equation}
where we indicated $\boldsymbol{\s} = (\s_{_{1}} \, , \ldots , \s_{n})$ and $\boldsymbol{\t} = (\t_{_{1}} \, , \ldots , \t_{n})\,$. The relations (\ref{Weyl CCR 1}) and (\ref{Weyl CCR 2}) are known as the \emph{Weyl form} of the CCR, or brief\mbox{}ly as the Weyl relations. The Schr\"{o}dinger representation of the Weyl relations is the one where $\hU$ and $\hV$ are generated by the operators (\ref{Schrodinger Q})--(\ref{Schrodinger P}).

\subsubsection{Unitary equivalence theorems}
\label{Unitary equivalence theorems}

A central problem was to establish under which hypotheses a representation of the CCR in a separable Hilbert space $\mscrH$ is unitarily equivalent to the Schr\"{o}dinger representation in $L^{2}(\mbbR^{n}, \mbbC)$. In other words, considering the case of $n$ degrees of freedom, under which conditions there exist a unitary map $\mcalU: \mscrH  \to L^{2}(\mbbR^{n}, \mbbC)$ such that
\begin{align}
\mcalU \hQ_{j} \, \mcalU^{-1} \, \ps(x_{_{1}} \, , \ldots , x_{n}) &= x_{j} \, \ps(x_{_{1}} \, , \ldots , x_{n}) \quad , \label{hQ map to L2} \\[2mm]
\mcalU \hP_{j} \, \mcalU^{-1} \, \ps(x_{_{1}} \, , \ldots , x_{n}) &= i \hbar \, \frac{\de}{\de x_{j}} \, \ps(x_{_{1}} \, , \ldots , x_{n}) \quad . \label{hP map to L2}
\end{align}
These representations are usually termed \emph{regular} representations of the CCR.

It is worthwhile to brief\mbox{}ly comment on the mathematical subtleties involved in these matters. It is possible to show from the abstract relations (\ref{CCR}) that there do not exist representations of the CCR in a f\mbox{}inite dimensional Hilbert space, and neither it is possible to represent them with bounded operators $\hQ_{j}$ and $\hP_{j}$ in the inf\mbox{}inite dimensional case \cite{Rosenberg 2004, Summers}. Since an unbounded operator cannot be def\mbox{}ined on the whole of $\mscrH$, the relations (\ref{CCR}) can at best hold in the common domain, dense in $\mscrH$, of the operators $\hQ_{j} \hP_{j}$ and $\hP_{j} \hQ_{j}$ for $j = 1 \, , \ldots , n\,$. The Weyl form of the CCR, on the other hand, may be considered to hold on the whole of $\mscrH$, since the exponentiated operators $\hU(\boldsymbol{\s})$ and $\hV(\boldsymbol{\t})$ are bounded and their domain can therefore be extended to the whole Hilbert space.

The best known result in this direction goes under the name of Stone-von Neumann theorem, which was formulated by Stone \cite{Stone 1930} with extremely concise hints of a possible proof, and proved by von Neumann \cite{von Neumann 1931}. It can be formulated \cite{Summers} as follows: let $\mscrH$ be a separable Hilbert space over $\mbbC$, and let the family of (weakly continuous) transformations $\ti{U}(\boldsymbol{\s})$ and $\ti{V}(\boldsymbol{\t})$, with $\boldsymbol{\s}$ and $\boldsymbol{\t} \in \mbbR^{n}$, be irreducible in $\mscrH$.\footnote{That is, the only subspaces of $\mscrH$ which are left invariant by $\ti{U}(\boldsymbol{\s})$ and $\ti{V}(\boldsymbol{\t})$ are the zero vector and $\mscrH$ itself.} If $\ti{U}$ and $\ti{V}$ satisfy the Weyl relations, then there exists a unitary transformation $\mcalU: \mscrH  \to L^{2}(\mbbR^{n}, \mbbC)$ such that 
\begin{align}
\mcalU \, \ti{U}(\boldsymbol{\s}) \, \mcalU^{-1} &= \hU(\boldsymbol{\s}) \quad , & \mcalU \, \ti{V}(\boldsymbol{\t}) \, \mcalU^{-1} &= \hV(\boldsymbol{\t}) \quad ,
\end{align}
where $\big\{ \hU , \hV \big\}$ is the Schr\"{o}dinger representation of the Weyl relations on $L^{2}(\mbbR^{n}, \mbbC)\,$. Moreover, if $\big\{ \ti{U} , \ti{V}, \mscrH \big\}$ is a reducible representation, then $\mscrH$ decomposes into a direct sum of countably many closed subspaces, on each of which the restriction of $\big\{ \ti{U} , \ti{V} \big\}$ is once again unitarily equivalent to the Schr\"{o}dinger representation.

Colloquially the theorem says that, in the irreducible case, under general conditions any representation of the Weyl form of the CCR is unitarily equivalent to the Schr\"{o}dinger representation, when the number of degrees of freedom is f\mbox{}inite. In this case the equations (\ref{hQ map to L2})--(\ref{hP map to L2}) between the generators of $\ti{U}$, $\ti{V}$, $\hU$ and $\hV$ then hold, so the associated representations of the (ordinary form of the) CCR are also unitarily equivalent. It is natural to wonder whether it is possible to prove the unitary equivalence of representations of the (ordinary) CCR with Schr\"{o}dinger's one, \emph{without} relying on the Weyl form of the CCR. This is indeed possible, as shown for example by Rellich \cite{Rellich 1946} and Dixmier \cite{Dixmier 1958}, by imposing additional conditions on the operators $\hQ$s and $\hP$s. A dif\mbox{}ferent, but related, question concerns the precise relation between the ordinary and the Weyl form of the CCR. In light of the technical complexity of these results, we direct to the exposition \cite{Putnam 1967} for a detailed treatment.

It is f\mbox{}inally important to mention that there exist representations of the (ordinary) CCR, still with a f\mbox{}inite number of degrees of freedom, which are \emph{not} unitarily equivalent to Schr\"{o}dinger's one. Of course, in these case the hypotheses of Rellich and Dixmier are not satisf\mbox{}ied, and the exponentiation of the position and momentum operators do not satisfy the Weyl form of the CCR. Far from being just mathematical curiosities, some of these representations are of common use in QM. A well-known example is that of a particle in a one-dimensional box with inf\mbox{}inite walls or with periodic boundary conditions.\footnote{In these cases the spectrum of the momentum operator is discrete, so the relevant representation of the CCR cannot be unitarily inequivalent to the Schr\"{o}dinger's one in $L^{2}(\mbbR, \mbbC)$, where the momentum operator has continuous spectrum.} A somehow more sophisticated example is that of QM in topologically non-trivial spaces, such as $\mbbR^{3}$ deprived of a line, which for instance is relevant to the Aharonov-Bohm ef\mbox{}fect.

\subsubsection{Application to the equivalence of theories}

A consequence of the unitary equivalence theorems is that, in an appropriate sense, every formulation of QM which satisf\mbox{}ies the hypotheses is equivalent to the Schr\"{o}dinger formulation.

Recall in fact that every observable quantity in QM is expressible in terms of expectation values, and consider a self-adjoint and irreducible representation $\big( \hQ_{j}, \hP_{j} \big)_{\!j = 1 \, , \ldots , n}$ of the CCR, and an analytic function $F\big[\big( \hQ_{j}, \hP_{j} \big)_{^{\!j}} \big]$ of these operators. Note that from the analyticity it follows that, if $\mcalU: \mscrH  \to L^{2}(\mbbR^{n}, \mbbC)$ is unitary, then we have
\begin{equation*}
F \Bigg[ \bigg( x_{j} \cdot , i \hbar \, \frac{\de}{\de x_{j}} \bigg)_{\!\!j} \Bigg] = F \bigg[ \Big( \mcalU \hQ_{j} \, \mcalU^{-1} , \mcalU \hP_{j} \, \mcalU^{-1} \Big)_{\!j} \bigg] = \mcalU F \Big[ \big( \hQ_{j}, \hP_{j} \big)_{\!j} \Big] \mcalU^{-1} \quad ,
\end{equation*}
and so
\begin{equation}
\Bigg\langle \mcalU \ps \, \Bigg\vert \, F \Bigg[ \bigg( x_{j} \cdot , i \hbar \, \frac{\de}{\de x_{j}} \bigg)_{\!\!j} \Bigg] \, \Bigg\vert \,\, \mcalU \ps \Bigg\rangle_{\!\!L^{2}(\mbbR^{n}, \mbbC)} = \bigg\langle \ps \, \bigg\vert \, F \Big[ \big( \hQ_{j}, \hP_{j} \big)_{\!j} \Big] \, \bigg\vert \, \ps \bigg\rangle_{\!\!\mcalH} \quad .
\end{equation}
If we set-up a correspondence between the states of $\mscrH$ and those of $L^{2}(\mbbR^{n}, \mbbC)$, physically identifying the states which are linked by the map $\mcalU$ (that is, we identify $\ps \in \mscrH$ with $\mcalU \ps \in L^{2}(\mbbR^{n}, \mbbC)$), then the expectation values are preserved by the correspondence between the two theories. If in addition the time evolution in the two theories commutes with this correspondence, then the two theories are physically equivalent (at least as far as analytic observables are concerned).

This is indeed the case of the Heisenberg-Born-Jordan matrix mechanics and of Schr\"{o}dinger's wave mechanics, whose physical equivalence can be proved resorting to the Stone-von Neumann theorem. More in general, the unitary equivalence theorems provide a precise mathematical basis for the assertion that the theory can be formulated abstractly, by specifying its algebraic structure, thereby substantiating Dirac's assertion.

\subsubsection{Application to the interaction picture}
\label{Application to the interaction picture}

These results also have bearing on the existence of the interaction picture. The argument is similar to that of the previous section, but here the two ``theories'' in question are in fact descriptions of the same physical system by switching on and of\mbox{}f an interaction (e.g.\ with a external potential).

Consider a physical system whose quantum dynamics is described, at abstract level, by a self-adjoint Hamiltonian $\hH_{_{0}} = h_{_{0}} \big[\big( \hQ_{j}, \hP_{j} \big)_{^{\!j}} \big]$ acting on a separable Hilbert space $\mscrH_{_{0}}$, where $h_{_{0}}$ is an analytic function and $\big( \hQ_{j}, \hP_{j} \big)_{\!j = 1 \, , \ldots , n}$ is an irreducible self-adjoint representation of the CCR. Consider another description of the quantum dynamics of the ``same'' physical system where an interaction is taken into account, so the self-adjoint Hamiltonian is
\begin{equation}
\hat{\mcalH} = \hat{\mcalH}_{_{0}} + \hat{\mcalH}_{_{\textup{int}}} \quad .
\end{equation}
with
\begin{align}
\hat{\mcalH}_{_{0}} &= h_{_{0}}\big[\big( \hq_{j}, \hp_{j} \big)_{^{\!j}} \big] \quad , & \hat{\mcalH}_{_{\textup{int}}} &= h_{_{\textup{int}}}\big[ \big( \hq_{j}, \hp_{j} \big)_{^{\!j}} \big] \quad .
\end{align}
Here $\hat{\mcalH}_{_{\textup{int}}}$ describes the interaction, with $h_{_{\textup{int}}}$ assumed to be analytic, and $\big( \hq_{j}, \hp_{j} \big)_{\!j = 1 \, , \dots \ n}$ is again an irreducible self-adjoint representation of the CCR. Let $\mscrH$ be the separable Hilbert space on which $\hat{\mcalH}$ acts. The question presents itself of how to relate the two descriptions, for then to introduce the interaction picture.

One option is to realize the two descriptions on a unique concrete space, $L^{2}(\mbbR^{n}, \mbbC)$, taking advantage of the unitary equivalence. Then $\hQ_{j}, \hP_{j}$ and $\hq_{j}, \hp_{j}$ are mapped to the same operators on $L^{2}(\mbbR^{n}, \mbbC)$, and this implies that the realization of the operator $\hH_{_{0}}$ coincides with that of $\hat{\mcalH}_{_{0}}\,$, which describe the free evolution of the system. One can then introduce the interaction picture as described above.

Another option is to work at the abstract level. This is especially compelling when it is possible to construct the space $\mscrH_{_{0}}$, together with the dynamics in it, purely by algebraic means (like when $h_{_{0}}$ describes a collection of free harmonic oscillators). The unitary equivalence of both the representations on $\mscrH_{_{0}}$ and on $\mscrH$ with the Schr\"{o}dinger representation on $L^{2}(\mbbR^{n}, \mbbC)$ implies that there exists a unitary map $\mcal{U} : \mscrH_{_{0}} \to \mscrH$ such that 
\begin{align}
\mcal{U} \, \hQ_{j} \, \mcal{U}^{-1} &= \hq_{j} \quad , & \mcal{U} \, \hP_{j} \, \mcal{U}^{-1} &= \hp_{j} \quad .
\end{align}
This implies f\mbox{}irst of all that
\begin{align}
\hat{\mcalH}_{_{0}} &= h_{_{0}} \big[ \big( \hq_{j}, \hp_{j} \big)_{^{\!j}} \big] = \, \mcalU \, h_{_{0}} \big[ \big( \hQ_{j}, \hP_{j} \big)_{^{\!j}} \big] \,\, \mcalU^{-1} = \mcalU \, \hH_{_{0}} \,\, \mcalU^{-1} \quad , \label{first of all 1} \\[2mm]
\hat{\mcalH}_{_{\textup{int}}} &= h_{_{\textup{int}}} \big[ \big( \hq_{j}, \hp_{j} \big)_{^{\!j}} \big] = \, \mcalU \, h_{_{\textup{int}}} \big[ \big( \hQ_{j}, \hP_{j} \big)_{^{\!j}} \big] \,\, \mcalU^{-1} = \mcalU \, \hH_{_{\textup{int}}} \,\, \mcalU^{-1} \quad , \label{first of all 2}
\end{align}
where we def\mbox{}ined $\hH_{_{\textup{int}}} = h_{_{\textup{int}}} \big[ \big( \hQ_{j}, \hP_{j} \big)_{^{\!j}} \big]\,$. It follows that $\hat{\mcalH}_{_{0}}$ is self-adjoint on $\mscrH$, and therefore so is $\hat{\mcalH}_{_{\textup{int}}} = \hat{\mcalH} - \hat{\mcalH}_{_{0}}$. Moreover, if $\ps(t)$ solves the Schr\"{o}dinger equation in $\mscrH_{_{0}}$ then $\f(t) = \mcalU \ps(t)$ solves in $\mscrH$ the equation
\begin{equation} \label{Setiba}
i \hbar \, \frac{\de}{\de t} \, \f(t) = \hat{\mcalH}_{_{0}} \, \f(t) \quad ,
\end{equation}
with the relation
\begin{equation} \label{Preserved}
\Big\langle \f(t) \, \Big\vert \, F \big[ \big( \hq_{j}, \hp_{j} \big)_{^{\!j}} \big] \, \Big\vert \, \f(t) \Big\rangle_{\!\mcalH} = \Big\langle \ps(t) \, \Big\vert \, F\big[\big( \hQ_{j}, \hP_{j} \big)_{^{\!j}} \big] \, \Big\vert \, \ps(t) \Big\rangle_{\!\mcalH_{_{0}}}
\end{equation}
being satisf\mbox{}ied for every analytic observable $F$. This means that, to any solution $\ps(t)$ of the (free) Schr\"{o}dinger equation in $\mscrH_{_{0}}$, we can associate (in a bijective way) a time-dependent conf\mbox{}iguration $\f(t) = \mcalU \ps(t)$ in $\mscrH$ which has the same physical properties (same expectation values of the observables) as $\ps(t)$. Since the time evolution of $\f(t)$, as (\ref{Setiba}) shows, is dictated by $\hat{\mcalH}_{_{0}}$, we can legitimately identify the latter as as the free Hamiltonian on $\mscrH$, and consequently introduce the interaction picture using $e^{\pm \frac{i}{\hbar}(t - t_{_{0}}) \hat{\mcalH}_{_{0}}}$.

Alternatively, we can map the interacting dynamics onto $\mscrH_{_{0}}$. Considering a solution $\f(t)$ of the Schr\"{o}dinger equation
\begin{equation}
i \hbar \, \frac{\de}{\de t} \, \f(t) = \hat{\mcalH} \, \f(t) \quad ,
\end{equation}
the relation
\begin{multline*}
\mcalU^{-1} \, \hat{\mcalH} \,\, \mcalU = \mcalU^{-1} \, h_{_{0}}\big[\big( \hq_{j}, \hp_{j} \big)_{^{\!j}} \big] \,\, \mcalU + \mcalU^{-1} \, h_{_{\textup{int}}}\big[ \big( \hq_{j}, \hp_{j} \big)_{^{\!j}} \big] \,\, \mcalU = \\[2mm]
= h_{_{0}} \big[ \big( \hQ_{j}, \hP_{j} \big)_{^{\!j}} \big] + h_{_{\textup{int}}} \big[ \big( \hQ_{j}, \hP_{j} \big)_{^{\!j}} \big] = \hH_{_{0}} + \hH_{_{\textup{int}}}
\end{multline*}
implies that, calling $\ps(t) = \mcalU^{-1} \f(t)$, we have
\begin{equation} \label{Setiba int}
i \hbar \, \frac{\de}{\de t} \, \ps(t) = \Big( \hH_{_{0}} + \hH_{_{\textup{int}}} \Big) \, \ps(t) \quad .
\end{equation}
Moreover, (\ref{Preserved}) again holds. This means that, to every solution $\f(t)$ of the (interacting) Schr\"{o}dinger equation in $\mscrH$ we can associate (in a bijective way) a time-dependent conf\mbox{}iguration $\ps(t) = \mcalU^{-1} \f(t)$ in $\mscrH_{_{0}}$ which has the same physical properties as $\f(t)$ (since (\ref{Preserved}) holds). Comparing (\ref{Setiba int}) with the (free) Schr\"{o}dinger equation in $\mscrH_{_{0}}$
\begin{equation} \label{Setiba free}
i \hbar \, \frac{\de}{\de t} \, \ps(t) = \hH_{_{0}} \, \ps(t) \quad ,
\end{equation}
it follows that we can meaningfully identify $\hH_{_{\textup{int}}}$ as the Hamiltonian on $\mscrH_{_{0}}$ which describes the interaction, and therefore introduce the interaction picture on $\mscrH_{_{0}}$ by means of $e^{\pm \frac{i}{\hbar}(t - t_{_{0}}) \hH_{_{0}}}$.

\subsubsection{Inf\mbox{}inite number of degrees of freedom}

It is a matter of fact that the Stone-von Neumann theorem does not hold when the number of degrees of freedom is inf\mbox{}inite, that is when the physical system under study is described by an inf\mbox{}inite collection of couples of canonical variables satisfying the CCR. This fact is relevant both for thermodynamic considerations in quantum mechanics, and for the quantization of f\mbox{}ield theories. Regarding the f\mbox{}irst case, it is a common procedure to consider the ``thermodynamic limit'' of a system by sending to inf\mbox{}inity both the number of particles $N$ and the volume $V$ in such a way to keep f\mbox{}ixed the density $N/V$. Regarding the second, it is well known that expanding a f\mbox{}ield $\f(t , x)$ over a set of orthonormal spatial functions one gets an equivalent description of the f\mbox{}ield in terms of the expansion coef\mbox{}f\mbox{}icients, which are functions of time and inf\mbox{}inite in number.

In these cases, the existence of representations of the CCR which are unitarily inequivalent is an unavoidable fact. While in (f\mbox{}inite volume) QM the irregular representations of the CCR (mentioned in the last paragraph of section \ref{Unitary equivalence theorems}) may be regarded as associated to peculiar situations, or in some cases circumvented by insisting on working on $L^{2}(\mbbR^{n}, \mbbC)$, in the quantum theory of f\mbox{}ields the existence of Unitarily Inequivalent Representations (UIR) of the CCR is the norm. The question then poses itself of how to select the appropriate representation for the specif\mbox{}ic system under study. For example: is it selected kinematically or dynamically? In the f\mbox{}irst case, which is the right criterion to choose it? In  the second, which is the dynamical mechanism which selects it?

We do not aim to answer this general questions here, and postpone further discussion on these matters to section \ref{sec: Comments}. Anyway, a brief comment is in order. Consider two representations of the CCR, $\big( \hQ_{j}, \hP_{j} \big)_{^{\!j}}$ and $\big( \hq_{j}, \hp_{j} \big)_{^{\!j}}$, which describe respectively the free and the interacting dynamics of a physical system with an inf\mbox{}inite number of degrees of freedom (so $j$ now belongs to an inf\mbox{}inite index set). Suppose that, whether kinematically or dynamically selected, they turn out to be unitarily inequivalent. Then, recalling the considerations of section \ref{Application to the interaction picture} about the interaction picture, a priori there is no way to claim that $\hat{\mcalH}_{_{0}}$ encodes the free dynamics on the space of the interacting theory, or that $\hH_{_{\textup{int}}}$ encodes the interaction on the space of the unperturbed theory. Furthermore, suppose we unilaterally decide to use unitarily equivalent representations in the two cases, and some inconsistency turns up when performing calculations in the interaction picture. Then we cannot really tell whether the problem is with the model we are using, or with the choice of representations, or both. Whether or not an incorrect choice of representations can explain the divergences which appear in perturbative QFT is one of the main subjects of the article.

\section{Van Hove's model and its stationary states}
\label{sec: stationary states}

Let us now turn to the study of van Hove's model, and consider a real, massive and relativistic scalar f\mbox{}ield which interacts with two (in the original formulation, $\scn$) point-like sources. The aim is to study the ef\mbox{}fect of the sources on the f\mbox{}ield, neglecting instead the backreaction of the f\mbox{}ield on the sources. Therefore, the system is idealized by assuming the sources to be inf\mbox{}initely massive, so that their positions $\vecy_{_{1}}$, $\vecy_{_{2}}$ (with $\vecy_{_{1}} \neq \vecy_{_{2}}$ strictly) remain f\mbox{}ixed. The interaction of the f\mbox{}ield with each source is modeled by a Hamiltonian which is simply proportional to the value of the scalar f\mbox{}ield at the position of the source, times a charge $q_{i}$ characteristic to each source.

\subsection{The two-sources case}
\label{subsec: two sources case}

The system is therefore described by the Hamiltonian
\begin{align}
H &= H_{_{0}} + g H_{_{\textup{int}}} \quad , \label{H tot} \\[5mm]
H_{_{0}} &= \half \, \int_{\mscr{P}} \! \Big( \pi^{2} + \babs{\vec{\nabla} \f}^{2} + m^{2} \f^{2} \Big) \, d^{3} x \quad , \label{H 0} \\[2mm]
H_{_{\textup{int}}} &= q_{_{1}} \, \f(t, \vecy_{_{1}}) + q_{_{2}} \, \f(t, \vecy_{_{2}}) \quad , \label{H i}
\end{align}
where $\f = \f(t,\vecx)$ is the real scalar f\mbox{}ield of mass $m\,$, $\pi = \pi(t,\vecx)$ the (real) conjugate momentum and $g$ is the coupling constant, supposed adimensional. Natural units are used, so that $\hbar = c = 1\,$. It is apparent that a suitable rescaling of $g$ and of $q_{_{1}}$ and $q_{_{2}}$ leaves the Hamiltonian unchanged. To avoid this arbitrariness, unless explicitly stated otherwise (notably sections \ref{sec: QFT and perturbative} and \ref{sec: S matrix and renormalization}) we work directly with the products $g_{_{1}} = g \, q_{_{1}}$ and $g_{_{2}} = g \, q_{_{2}}\,$. The Hamiltonian $H$ therefore depends on the parameters $g_{_{1}}$, $g_{_{2}}$, $\vecy_{_{1}}$ and $\vecy_{_{2}}$.

The system is supposed to be enclosed in a cubic box of large volume, with periodic boundary conditions.\footnote{Alternatively, the system can be supposed to live on a f\mbox{}lat $3$-torus of the same volume.} In other words, the f\mbox{}ield and the conjugate momentum are supposed to belong to the space of square-integrable functions periodic on a cubic lattice, where $L$ indicates the length of the edges of the primitive cell $\mscr{P}$ and $V = L^{3}$ indicates its volume. The reciprocal lattice, which we indicate with $\mcal{R}\,$, is again a cubic lattice whose primitive cell has edges of length $2 \pi/ L$ and volume equal to $(2 \pi)^{3}/ V\,$. Performing a Fourier expansion we obtain
\begin{align}
\f(t, \vecx) &= \frac{1}{\sqrt{2 V}} \, \sum_{\veck \in \mcalR} \, \frac{1}{\sqrt{\o_{k}}} \,\, \bigg( a^{\phantom{\ast}}_{^{\veck}} \,\, e^{-i \, \o_{k}^{\phantom{1}} t} \, e^{i \veck \cdot \vecx} + a^{\ast}_{^{\veck}} \,\, e^{i \, \o_{k}^{\phantom{1}} t} \, e^{- i \veck \cdot \vecx} \bigg) \\[2mm]
\pi(t, \vecx) &= \frac{-i}{\sqrt{2 V}} \, \sum_{\veck \in \mcalR} \, \sqrt{\o_{k}} \,\, \bigg( a^{\phantom{\ast}}_{^{\veck}} \,\, e^{-i \, \o_{k}^{\phantom{1}} t} \, e^{i \veck \cdot \vecx} - a^{\ast}_{^{\veck}} \,\, e^{i \, \o_{k}^{\phantom{1}} t} \, e^{- i \veck \cdot \vecx} \bigg) \quad ,
\end{align}
where $\o_{k}^{\phantom{1}} = \sqrt{m^{2} + k^{2\,}}$ and $k = \bnorm{\veck}\,$. The wavevectors belonging to $\mcalR$ are of the form $\veck = \vec{n} \,\, 2 \pi/L$ where $\vec{n} \in \mbbZ^{3}$.

Before turning to the quantization of this system, it is useful to mention a procedure which will be useful below, the inf\mbox{}inite volume limit. By this we mean the limit where $L$ (and therefore $V$) is sent to inf\mbox{}inity in such a way to keep the basis functions normalized, which is implemented by transforming the sum over the reciprocal lattice into an integral as follows
\begin{equation}
\frac{1}{V} \, \sum_{\veck \in \mcalR} \,\, \to \,\, \frac{1}{(2 \pi)^{3}} \int_{\mbbR^{3}} d^{3} k \quad .
\end{equation}

\subsubsection{Canonical quantization}

The canonical quantization of the system is performed in the Schr\"{o}dinger picture. As usual, the Fourier coef\mbox{}f\mbox{}icients are promoted to operators which obey the commutation relations
\begin{align} \label{a adg commutation relations}
\Big[ \, \ha^{\phantom{\dagger}}_{^{\veck}} \, , \ha^{\phantom{\dagger}}_{^{\veck^{\p}}} \Big] = \Big[ \, \ha^{\dagger}_{^{\veck}} \, , \ha^{\dagger}_{^{\veck^{\p}}} \Big] &= 0 & \Big[ \ha^{\phantom{\dagger}}_{^{\veck}} \, , \ha^{\dagger}_{^{\veck^{\p}}} \Big] &= \d_{\veck \, , \, \veck^{\p}} \quad ,
\end{align}
so the f\mbox{}ield and the conjugate momentum become the (time-independent) operators
\begin{align}
\hf(\vecx) &= \frac{1}{\sqrt{2V}} \, \sum_{\veck \in \mcalR} \frac{1}{\sqrt{\o_{k}}} \, \Big( \ha^{\phantom{\dagger}}_{^{\veck}} \, e^{i \veck \cdot \vecx} + \ha^{\dagger}_{^{\veck}} \, e^{- i \veck \cdot \vecx} \Big) \label{quantised phi} \\[2mm]
\hpi(\vecx) &= \frac{-i}{\sqrt{2V}} \, \sum_{\veck \in \mcalR} \sqrt{\o_{k}} \, \Big( \ha^{\phantom{\dagger}}_{^{\veck}} \, e^{i \veck \cdot \vecx} - \ha^{\dagger}_{^{\veck}} \, e^{- i \veck \cdot \vecx} \Big) \label{quantised pi} \quad ,
\end{align}
which obey the canonical commutation relations (for f\mbox{}ields) $\big[ \hf(\vecx) , \hpi(\vecy) \big] = i \, \d(\vecx - \vecy)\,$. The Hamiltonian operator, neglecting as usual in $\hH_{_{0}}$ the (inf\mbox{}inite) ground state energy of the oscillators, takes the form
\begin{align}
\hH_{_{0}} &= \sum_{\veck \in \mcalR} \o_{k}^{\phantom{1}} \,\, \ha^{\dagger}_{^{\veck}} \, \ha^{\phantom{\dagger}}_{^{\veck}} \quad , \nn \\[1mm]
g \hH_{_{\textup{int}}} &= \sum_{i = 1}^{2} \frac{g_{_{i}}}{\sqrt{2V}} \, \sum_{\veck \in \mcalR} \, \frac{1}{\sqrt{\o_{k}}} \, \Big( \ha^{\phantom{\dagger}}_{^{\veck}} \, e^{i \veck \cdot \vecy_{i}} + \ha^{\dagger}_{^{\veck}} \, e^{- i \veck \cdot \vecy_{i}} \Big) \quad . \label{HI HB a a*}
\end{align}

At this point the article departs from the ``standard'' exposition of the canonical quantization of a real scalar f\mbox{}ield given in introductory textbooks on QFT. This is done by introducing the families of Hermitian operators
\begin{align} \label{qk pk}
\hq_{_{\veck}} &= \frac{1}{\sqrt{2}} \, \Big( \ha^{\phantom{*}}_{^{\veck}} +  \ha^{\dg}_{^{\veck}} \Big) & \hp_{_{\veck}} &= \frac{i}{\sqrt{2}} \, \Big( \ha^{\dagger}_{^{\veck}} - \ha^{\phantom{*}}_{^{\veck}} \Big) \quad ,
\end{align}
which satisfy the canonical commutation relations
\begin{align} \label{q p CR}
\Big[ \, \hq_{_{\veck}} \, , \hq_{_{\veck^{\p}}} \Big] &= \Big[ \, \hp_{_{\veck}} \, , \hp_{_{\veck^{\p}}} \Big] = 0 \quad , & \Big[ \, \hq_{_{\veck}} \, , \hp_{_{\veck^{\p}}} \Big] &= i \, \d_{\veck \, , \, \veck^{\p}} \quad .
\end{align}
In terms of the new operators we have
\begin{equation} \label{Hamiltonian tau omega -1}
\hH = \frac{1}{2} \, \sum_{\veck \in \mcalR} \, \o_{k}^{\phantom{1}} \, \Big( \hp^{\,2}_{_{\!\veck}} + 2 \, \s_{_{\!\veck}} \, \hp_{_{\veck}} + \hq^{\,2}_{_{\veck}} + 2 \, \t_{_{\!\veck}} \, \hq_{_{\veck}} - 1 \Big) \quad ,
\end{equation}
where
\begin{equation} \label{tau e omega}
\t_{_{\!\veck}} = \frac{1}{\sqrt{V \o_{k}^{3}}} \, \sum_{i = 1}^{2} \, g_{_{i}} \cos \big( \veck \cdot \vecy_{i} \big) \,\,\,\, , \,\,\,\, \s_{_{\!\veck}} = - \frac{1}{\sqrt{V \o_{k}^{3}}} \, \sum_{i = 1}^{2} \, g_{_{i}} \sin \big( \veck \cdot \vecy_{i} \big) \quad .
\end{equation}
Completing the squares the Hamiltonian operator becomes
\beq \label{total Hamiltonian}
\hH = \frac{1}{2} \, \sum_{\veck \in \mcalR} \, \o_{k}^{\phantom{1}} \, \bigg[ \, \Big( \hp_{_{\!\veck}} + \s_{_{\!\veck}} \Big)^{\! 2} + \Big( \hq_{_{\veck}} + \t_{_{\!\veck}} \Big)^{\! 2} - 1 \, \bigg] + B + C \quad ,
\eeq
where
\begin{align}
B &= - \frac{g_{_{1}} g_{_{2}}}{V} \, \sum_{\veck \in \mcalR} \, \frac{\cos \big[ \veck \cdot (\vecy_{_{1}} - \vecy_{_{2}}) \big]}{\o_{k}^{2}} \quad , \label{Yukawa int energy} \\[2mm]
C &= - \frac{g^{2}_{_{1}} + g^{2}_{_{2}}}{2 V} \, \sum_{\veck \in \mcalR} \, \frac{1}{\o_{k}^{2}} \label{Yukawa self-energy} \quad .
\end{align}

The number $B$ has a transparent physical interpretation. In the inf\mbox{}inite volume limit it can be computed explicitly giving
\begin{equation}
B = - \frac{g_{_{1}} g_{_{2}}}{4 \pi} \, \frac{e^{-m \, \norm{\vecy_{_{1}} - \vecy_{_{2}}}}}{\norm{\vecy_{_{1}} - \vecy_{_{2}}}} \quad ,
\end{equation}
which is the (Yukawa-type) interaction energy of the two sources. This interpretation readily extends to the case $\scn > 2\,$, where $B$ becomes the sum of the pair-wise interaction energy of the sources via the Yukawa potential mediated by $\f\,$. The number $C$ is instead divergent, and following van Hove we discard it hereafter (with the exception of section \ref{sec: S matrix and renormalization} where it will resurface). We aim to analyze its meaning and signif\mbox{}icance in a future publication.

\subsubsection{Representation of the commutation relations}
\label{subsubsec: Representation}

The article then chooses a representation of the commutation relations (\ref{q p CR}) in which the operators $\hq_{_{\veck}}$ are diagonal. More specif\mbox{}ically, let us indicate with $\mscrR$ the set of maps $\mcalR \to \mbbR$, which associate a real number to each point of the reciprocal lattice. The elements of $\mscrR$ can equivalently be thought as sequences of real numbers indexed by $\veck \in \mcalR\,$. To avoid cumbersome expressions, in the following the sequence $\{q_{_{\veck}}\}_{_{\!\veck \in \mcalR}}$ will be indicated simply with $\{ q \}\,$. The abstract state vector $\mid \! \Phi (t) \, \rangle$ of the f\mbox{}ield is then represented by a (time-dependent) functional
\begin{align} \label{functional}
\Phi &: \, \mbbR \times \mscrR \to \mbbR & \Phi = \Phi \big( t , \{ q \} \big) \quad ,
\end{align}
and the abstract operators $\hq_{_{\veck}}$ and $\hp_{_{\!\veck}}$ are realized as the operators $\hmcalQ_{_{\veck}}$ and $\hmcalP_{_{\!\!\veck}}$ acting on the space of functionals as follows
\begin{align} \label{differential}
\hmcalQ_{_{\veck}} \, \Phi &= q_{_{\veck}} \, \Phi & \hmcalP_{_{\!\!\veck}} \, \Phi &= - i \, \frac{\de}{\de q_{_{\veck}}} \, \Phi \quad .
\end{align}
Accordingly, the Hamiltonian operator (\ref{total Hamiltonian}) is realized as the operator $\hmcalH$ acting as
\beq \label{mathcal H}
\hmcalH \,\, \Phi = \Bigg[ \, B + \frac{1}{2} \, \sum_{\veck \in \mcalR} \, \o_{k}^{\phantom{1}} \, \Bigg( \! \bigg( \! - i \frac{\de}{\de q_{_{\veck}}} + \s_{_{\!\veck}} \bigg)^{\!\! 2} + \Big( q_{_{\veck}} + \t_{_{\!\veck}} \Big)^{\! 2} - 1 \Bigg) \, \Bigg] \, \Phi \quad .
\eeq
 
Since in $\hmcalH$ the operators relative to dif\mbox{}ferent $\veck$ do not mix, it is possible to f\mbox{}ind its eigenfunctionals by separation of variables. Indeed each operator in the inf\mbox{}inite sum on the right hand side of (\ref{mathcal H}) is closely related to the Hamiltonian of the (one dimensional) quantum harmonic oscillator, and indicating
\begin{equation} \label{varphi}
\vf_{_{\! n}}^{_{\, \veck}} \big( q_{_{\veck}} \big) = e^{-i \, \s_{_{\! \veck}} \, q_{_{\veck}}} \,\, \ps_{n} \Big( q_{_{\veck}} + \t_{_{\veck}} \Big)
\end{equation}
we have
\begin{equation}
\frac{1}{2} \, \Bigg[ \bigg( \! - i \frac{\de}{\de q_{_{\veck}}} + \s_{_{\!\veck}} \bigg)^{\!\! 2} + \Big( q_{_{\veck}} + \t_{_{\!\veck}} \Big)^{\! 2} - 1 \, \Bigg] \, \vf_{_{\! n}}^{_{\, \veck}} \big( q_{_{\veck}} \big) = n  \,\, \vf_{_{\! n}}^{_{\, \veck}} \big( q_{_{\veck}} \big) \quad ,
\end{equation}
where $n$ is a non-negative integer and $\ps_{n}$ is the standard Hermite function\footnote{The Hermite polynomials $H_{n}$ are def\mbox{}ined as $H_{n}(x) = (-1)^{n} \, e^{x^{2}} \frac{d^{n}}{dx^{n}} \, e^{-x^{2}}$.} of $n$-th degree
\beq \label{Hermite}
\ps_{n}\big( x \big) = \frac{1}{\sqrt{2^{n} \, n! \, \sqrt{\pi}}} \,\, e^{- \frac{x^{2}}{2}} \, H_{n}(x) \quad .
\eeq
The eigenfunctionals of $\hmcalH$ are therefore the inf\mbox{}inite products of functions
\beq
\label{eigenfunctions}
\Phi_{\{ n \}}  \big( \{ q \} \big) = \prod_{\veck \in \mcalR} \vf_{_{\! n(\veck)}}^{_{\, \veck}} \big( q_{_{\veck}} \big) \quad ,
\eeq
and their eigenvalues are
\beq
\label{spectrum}
\hmcalH \, \Phi_{\{ n \}} = \Bigg[ B + \sum_{\veck \in \mcalR} \, n(\veck) \,\, \o_{k}^{\phantom{1}} \, \Bigg] \, \Phi_{\{ n \}} \quad .
\eeq
Summing up, the (time-independent) eigenfunctionals of $\hmcalH$ are factorized into the real functions $\vf_{_{\! n(\veck)}}^{_{\, \veck}}$ of real variable $q_{_{\veck}}\,$, each associated to one vector of the reciprocal lattice. They are characterized by an integer ``excitation number'' $n(\veck) \geq 0\,$, a real ``phase factor'' $\s_{_{\! \veck}}$ and a real ``shift factor'' $\t_{_{\! \veck}}\,$, the dependence on the parameters $g_{_{1}}$, $g_{_{2}}$, $\vecy_{_{1}}$, $\vecy_{_{2}}$ of the model being entirely contained into the phase and shift factors. The eigenfunctionals $\Phi_{\{ n \}}$ are labeled by the collection $\{ n \} = \{n(\veck)\}_{_{\!\veck \in \mcalR}}$ of excitation numbers.

The expression (\ref{spectrum}) suggests a particle interpretation in which $n(\veck)$ is the number of quanta associated to the wavevector $\veck\,$, each of which has energy $\o_{k}^{\phantom{1}}\,$, present in the state $\Phi_{\{ n \}}\,$. Note that $\veck$ can be identif\mbox{}ied with the momentum of each quantum only asymptotically (i.e.\ spatially far away from the point-like sources), since the states (\ref{eigenfunctions}) are not eigenfunctions of the momentum operator. The ground state of the system is obtained by setting to zero all the $n(\veck)\,$, in which case the energy eigenvalue reduces to the potential energy of the sources. The only excited states with f\mbox{}inite energy are those for which just a f\mbox{}inite number of the $n(\veck)$ are non-vanishing.

\section{Space of the state vectors for the interacting f\mbox{}ield}
\label{sec: space of the state vectors}

With the result of the previous section in hand, the article proceeds to def\mbox{}ine the concept of ``space of the state vectors'' for the model under consideration, and to study what happens when the values of the parameters of the model are changed.

\subsection{The space of the state vectors}
\label{subsec: space of the state vectors}

As it is well-known, in ordinary Quantum Mechanics (i.e.\ with a f\mbox{}inite number of degrees of freedom), the possible physical states of the system are represented by the unit rays of a separable Hilbert space: the \emph{space of the state vectors}. The stationary states which correspond to all the possible values of the energy constitute an orthogonal (or orthogonalizable, in the case of degeneracy) basis of the space of the state vectors (focusing for simplicity on the case of discrete spectrum).

For the system under consideration here, a dif\mbox{}ference emerges already at this point. Due to the number of degrees of freedom being inf\mbox{}inite, there exist states characterized by f\mbox{}inite excitation numbers whose energy is nevertheless inf\mbox{}inite, to wit those states where an inf\mbox{}inite number of $n(\veck)$ are non-vanishing. Since these states cannot be excited from the ground state by any physical process, because this would require an inf\mbox{}inite energy, the article suggests to adopt as the space of state vectors the Hilbert space generated by the stationary states of f\mbox{}inite energy, that is by the $\Phi_{\{ n \}}$ for which the non-zero $n(\veck)$ are in a f\mbox{}inite number. These basis vectors are normalized and pairwise orthogonal, and they are of the ``inf\mbox{}inite product'' form (\ref{eigenfunctions}) with the additional property that only a f\mbox{}inite number of factors $\vf_{_{\! n(\veck)}}^{_{\, \veck}}$ are dif\mbox{}ferent from $\vf_{_{\! 0 \phantom{\veck}}}^{_{\, \veck}}$.

To describe the main result of the article regarding the dependence of the space of the state vectors on the parameters of the model, it is advisable to quote directly van Hove's words (his emphasis):
\begin{quotation}
For the usual systems with a f\mbox{}inite number of degrees of freedom, such as those made of particles under the inf\mbox{}luence of external forces, a change in the constants which appear into the Hamiltonian (charges, constants which determine the external potential, etc.) produces a change in the stationary states but does not change the space of the state vectors. The latter remains the same, the new stationary states being linear combinations of the initial ones. The essential point we wish to emphasize is that this is not the case anymore for the bosonic f\mbox{}ield studied above. \emph{Any variation of the constants $\scn$, $g_{_{i}}$, $\vecy_{_{i}}$ replaces the space of state vectors of the f\mbox{}ield with another one which does not have any (non-null) vector in common with the former. Moreover, any vector of the f\mbox{}irst space has vanishing scalar product with any vector of the second space.}
\end{quotation}
These words of course refer to the generic case with $\scn$ sources. It seems to us that f\mbox{}ixing $\scn = 2$ does not weaken the relevance of the result, since the phenomenon of ``replacement of the space of state vectors'' happens with any change of the remaining parameters $g_{_{i}}$, $\vecy_{_{i}}$ and is therefore still well represented.

\subsection{Dependence of the space on the parameters}
\label{subsec: dependence on the parameters}

Let us now investigate how the space of state vectors depends on the parameters of the model.

Be $\G = \big\{ (g_{_{1}}, \vecy_{_{1}})\, , (g_{_{2}}, \vecy_{_{2}}) \big\}$ and $\barG = \big\{ (\bar{g}_{_{1}}, \vecz_{_{1}}) \, , (\bar{g}_{_{2}} , \vecz_{_{2}}) \big\}$ two choices for the parameters, with $\vecy_{_{1}} \neq \vecy_{_{2}}$ and $\vecz_{_{1}} \neq \vecz_{_{2}}$. For the sake of brevity let us agree that, when we say ``the case $\G$'', we mean ``the case where the value of the parameters is $\G$'', and an analogous meaning is attached to ``the states of $\G$'' and similar expressions. Let us also agree that the notation $\sum n(\veck) < + \infty$ means that $n(\veck) \neq 0$ only for f\mbox{}inite set of values of $\veck\,$, and let us indicate with $\mscrS_{_{\!\!\G}}$ (respectively, $\mscrS_{_{\!\!\barG}}$) the space of state vectors of $\G$ (respectively, the space of state vectors of of $\barG$). The f\mbox{}inite energy stationary states of $\G$ are then the vectors $\Phi_{\{ n \}}$ of the form (\ref{eigenfunctions}) with $\sum n(\veck) < + \infty\,$, while those of $\barG$ have the analogous form
\beq \label{eigenfunctions bar}
\bar{\Phi}_{\{ \barn \}} \big( \{ q \} \big) = \prod_{\veck \in \mcalR} \bar{\vf}_{_{\! \barn(\veck)}}^{_{\, \veck}} \big( q_{_{\veck}} \big) \quad ,
\eeq
again with $\sum \barn(\veck) < + \infty\,$. In analogy to \eqref{varphi} and \eqref{tau e omega} the functions $\bar{\vf}$ are def\mbox{}ined as
\begin{equation} \label{barvarphi}
\bar{\vf}_{_{\! \barn\phantom{(\veck)}\!}}^{_{\, \veck}} \!\!\!\! \big( q_{_{\veck}} \big) = e^{-i \, \bars_{_{\! \veck}} \, q_{_{\veck}}} \,\, \ps_{\barn} \Big( q_{_{\veck}} + \bart_{_{\veck}} \Big) \quad ,
\end{equation}
where
\begin{equation} \label{tau bar e omega bar}
\bart_{_{\!\veck}} = \frac{1}{\sqrt{V \o_{k}^{3}}} \, \sum_{i = 1}^{2} \, \barg_{_{i}} \cos \big( \veck \cdot \vecz_{i} \big) \quad , \quad \bars_{_{\!\veck}} = - \frac{1}{\sqrt{V \o_{k}^{3}}} \, \sum_{i = 1}^{2} \, \barg_{_{i}} \sin \big( \veck \cdot \vecz_{i} \big) \quad .
\end{equation}
It is not dif\mbox{}f\mbox{}icult to see from (\ref{tau e omega}) and (\ref{tau bar e omega bar}) that $\G = \barG$ if and only if $\s_{_{\!\veck}} = \bars_{_{\!\veck}}$ and $\t_{_{\!\veck}} = \bart_{_{\!\veck}}$ for every $\veck \in \mcalR\,$.

To compare the space of state vectors of $\G$ and $\barG\,$, and in particular to speak of their orthogonality, it is necessary to specify how the scalar product between these states is def\mbox{}ined. In the article this is achieved by regarding the spaces $\mscrS_{_{\!\!\G}}$ and $\mscrS_{_{\!\!\barG}}$ as subsets of a larger Hilbert space $\mscrS$, using von Neumann's construction of the ``inf\mbox{}inite direct product'' spaces \cite{von Neumann Comp Math 1939}. Starting from a sequence of Hilbert spaces $\{ \mscrH_{j}\}_{j}$, von Neumann showed how the inf\mbox{}inite direct product of this spaces can be def\mbox{}ined, resulting into a new Hilbert space. This formalism can be readily applied to our case since the eigenstates (\ref{eigenfunctions}) and (\ref{eigenfunctions bar}) have the structure of inf\mbox{}inite products of functions. Calling $\mscrH$ the Hilbert space to which the functions belong, Von Neumann's construction then leads to a non-separable Hilbert space $\mscrS$ generated by the inf\mbox{}inite products of functions, with the scalar product in $\mscrS$ being def\mbox{}ined via the scalar product in $\mscrH$. More specif\mbox{}ically, indicating with $\prod f_{_{\!\veck}}$ and $\prod \bar{f}_{_{\!\veck}}$ two elements of $\mscrS$, in case the inf\mbox{}inite (numerical) product of the scalar products of the constituent functions converge we def\mbox{}ine
\begin{equation*}
\Bigg\langle \prod_{\veck \in \mcalR} f_{_{\!\veck}} \, \Bigg\vert \, \prod_{\veck \in \mcalR} \bar{f}_{_{\!\veck}} \Bigg\rangle_{\!\!\!\mscrS} = \prod_{\veck \in \mcalR} \Big\langle f_{_{\!\veck}} \, \Big\vert \, \bar{f}_{_{\!\veck}} \Big\rangle_{\!\!\mscrH} \quad .
\end{equation*}
Here the scalar product on the left hand side is that of $\mscrS$ while the scalar product on the right hand side is that of $\mscrH$. We omit hereafter the subscript indicating the space in which the scalar product is def\mbox{}ined. When the inf\mbox{}inite (numerical) product on the right-hand side does \emph{not} converge, the value of the scalar product on the left-hand side have to be assigned by convention, the simplest convention being to assign to it the value zero. The space $\mscrS$ is uncountably inf\mbox{}inite dimensional. It is not dif\mbox{}f\mbox{}icult to see that, with this def\mbox{}inition, the family of states $\Phi_{\{ n \}}$ (respectively, $\bar{\Phi}_{\{ n \}}$) with $\sum n(\veck) < + \infty$ constitute an orthonormal basis for $\mscrS_{_{\!\!\G}}$ (respectively, $\mscrS_{_{\!\!\barG}}$).

With these def\mbox{}initions at hand we can prove the following results (whose proofs we relegate to the appendix \ref{App: Orthogonality}, to improve the readability):
\begin{enumerate}
 \item if $\G \neq \barG\,$, \emph{a f\mbox{}inite energy stationary state of $\G$ and a f\mbox{}inite energy stationary state of $\barG$ always have vanishing scalar product};
 \item If $\G \neq \barG\,$, \emph{every f\mbox{}inite energy stationary state of $\G$ is orthogonal to every stationary state of $\barG\,$, independently of the latter having f\mbox{}inite or inf\mbox{}inite energy}.
\end{enumerate}
Clearly, the second result is more general of the f\mbox{}irst and includes it. The f\mbox{}irst result implies that, when $\G \neq \barG\,$, the scalar product between any two stationary states of $\mscrS_{_{\!\! \G}}$ and $\mscrS_{_{\!\! \barG}}$ vanishes, and the spaces $\mscrS_{_{\!\! \G}}$ and $\mscrS_{_{\!\! \barG}}$ constitute orthogonal subspaces of $\mscrS \!$, not having any element in common apart from the zero vector. The second result on the other hand has the following consequence, which is perhaps physically more intuitive. Again assuming $\G \neq \barG\,$, let us choose a sequence $\{n(\veck)\}_{_{\! \veck}}$ and, for every $\veck \in \mcalR\,$, expand $\vf_{_{\! n(\veck)}}^{_{\, \veck}}$ over the family $\{ \bar{\vf}_{_{0}}^{_{\, \veck}}$, $\bar{\vf}_{_{1}}^{_{\, \veck}}$, \ldots $\}\,$. Inserting this expansion into (\ref{eigenfunctions}) and taking into account (\ref{eigenfunctions bar}), we obtain an expansion of the state $\Phi_{\{ n \}}$ over the collection of states $\bar{\Phi}_{\{ \barn \}}$ where $\{ \barn \}$ varies over all conf\mbox{}igurations (that is, also over the conf\mbox{}igurations $\{ \barn \}$ with an inf\mbox{}inite number of non-vanishing components). The second result means that, when $\sum n(\veck) < + \infty\,$, all the coef\mbox{}f\mbox{}icients of this expansion vanish. Since $\Phi_{\{ n \}} \neq 0\,$, \emph{the f\mbox{}inite-energy stationary states of $\G$ cannot be expanded in series over the stationary states of $\barG\,$}. In particular, the f\mbox{}inite-energy stationary states of the f\mbox{}ield for $g \neq 0$ are not linear combinations of the stationary states of the free f\mbox{}ield.

\section{QFT and the perturbative approach}
\label{sec: QFT and perturbative}

The results obtained above continue to hold in the general case $\scn \geq 1\,$: any change in the number of sources $\scn\,$, in the constants $g_{_{i}}$ and in the position of the sources $\vecy_{_{i}}$, however small the change may be, changes the space of the state vectors into another space which is orthogonal to the previous one. It is important to understand the implications of this result for a realistic Quantum Field Theory, in light of the fact that the present model can be regarded at best as a toy model. Note that below we switch back to considering separately the coupling constant $g$ and the charges $q_{i}$ (in the stead of the products $g_{i} = g \, q_{i}$).

\subsection{Discussion on the generality of the result}
\label{subsec: generality of the result}

Van Hove is of the opinion that ``it is is totally possible that this type of situation, which does not exist for systems with a f\mbox{}inite number of degrees of freedom, is unavoidable for interacting quantum f\mbox{}ields''. He acknowledges that the present description of the interaction of the quantum f\mbox{}ield with the sources is very crude, and therefore that the variation of the space of the state vectors with the positions $\vecy_{_{i}}$ and the charges $q_{_{i}}$ of the sources should disappear in a theory which takes into account the reaction of the f\mbox{}ield on the sources (recoil ef\mbox{}fect). On the other hand, he puts forward the conjecture that the variation of the space with $\scn$ and $g$ may remain valid in full generality and constitute an important trait of the theory, not necessarily so from the physical point of view but especially from the formal point of view. In other words, he conjectures that also with a more ref\mbox{}ined description of the interaction of the f\mbox{}ield with the sources, the spaces $\mscrS_{\! g,\scn}$ and $\mscrS_{\barg, \bar{\scn}}$ would be notwithstanding orthogonal or at least dif\mbox{}ferent whenever $\scn \neq \bar{\scn}$ and/or $g \neq \barg\,$.

This conjecture is in fact supported by the results of his previous work \cite{van Hove 1951}, which considers the cases of electrodynamics and of the Yukawa theory of mesons, under the only hypothesis that f\mbox{}inite-energy stationary states exist in presence of interaction. It is indeed shown there that the total Hamiltonian, interaction included, cannot be considered as an operator acting on the space $\mscrS_{\! 0, \scn}$ describing the free f\mbox{}ields, since it does not map any vector of $\mscrS_{\! 0, \scn}$ into a vector of $\mscrS_{\!0, \scn}\,$. This fact is not changed by the addition of an inf\mbox{}inite constant such as (\ref{Yukawa self-energy}) to the Hamiltonian, as an analysis of the proof shows. It follows that the space $\mscrS_{\! 0, \scn}$ cannot contain any stationary state of the interacting f\mbox{}ields, and that, when the coupling constant $g$ is dif\mbox{}ferent from zero, the spaces $\mscrS_{\! 0, \scn}$ and $\mscrS_{\!\! g, \scn}$ are distinct. Van Hove further conjectures that, in analogy with the special case studied above, we may expect $\mscrS_{\! 0, \scn}$ and $\mscrS_{\!\! g, \scn}$ to be orthogonal subspaces of a ``inf\mbox{}inite direct product'' space $\mscrS_{\! \scn}$ constructed for the two f\mbox{}ields in a similar way to how the space $\mscrS$ was constructed in the previous section.

\subsection{Applicability of the perturbative approach}
\label{subsec: perturbative approach}

The formal results of the previous sections suggest profound implications regarding the applicability of the perturbative method in Quantum Field Theory. Again, it is best to quote directly van Hove's words (his emphasis):
\begin{quotation}
In light of the results above, it is interesting to remark how the perturbative approach seems to be badly suited to deal with interacting quantum f\mbox{}ields. Apart from the hypothesis of the interaction being weak, which seems justif\mbox{}ied at least when $g$ is small, this method assumes a priori that the ef\mbox{}fect of the interaction is to displace the f\mbox{}inite-energy stationary states of the free f\mbox{}ields \emph{inside the space $\mscrS_{\! 0, \scn}$ generated by them}. This second hypothesis, automatically satisf\mbox{}ied in the problems usually dealt with the perturbative approach, is frequently taken for granted. It is essential to underline that \emph{this second hypothesis is not satisf\mbox{}ied in the case of interacting quantum f\mbox{}ields, independently of how small the coupling constant be}. For the latter f\mbox{}ields, the interaction displaces the stationary states out of the space $\mscrS_{\! 0, \scn}$ moving them to a dif\mbox{}ferent space, $\mscrS_{\!\! g, \scn}\,$.

The considerations above clearly permit to understand the reason why the perturbative approach inevitably leads to mathematical problems, represented by the appearance of divergences.
\end{quotation}
It is worthwhile to highlight that, beyond questioning the validity of the perturbative approach in QFT, van Hove explicitly suggests a link between the appearance of perturbative divergences and the orthogonality of the free and the interacting spaces of state vectors.

Of course, as soon as the validity of the perturbative expansion in Quantum Field Theory is criticized, one is almost inevitably confronted with the problem of explaining why, when used together with the renormalization procedure, this expansion is so successful in describing the experimental results. It is useful to recall that the empirical success of the renormalized perturbative expansion in Quantum ElectroDynamics (QED) was fully appreciated when van Hove was working on the article, since for example the calculation of the Lamb shift \cite{Bethe 1947} and of the radiative corrections to the magnetic moment of the electron \cite{Schwinger 1948} were known since several years. In the article, van Hove does not tackle this issue in the context of QED, limiting to declare that ``surely this success will not be completely understood until a deeper comprehension is obtained on the stationary states in the context of electrodynamics''.

He remarks that the mathematical problems can be circumvented by placing a cut-of\mbox{}f on the wavevectors (i.e.\ on the momenta), that is considering only the Fourier modes with $\abs{\veck} \leq \La\,$. If we consider the model def\mbox{}ined by (\ref{HI HB a a*}) where the range of the sum over $\veck$ is taken to be $\mcalR_{_{\La}} = \{ \veck \in \mcalR \, , \, \abs{\veck} \leq \La \}\,$, we are ef\mbox{}fectively dealing with a system with a f\mbox{}inite number of degrees of freedom and we can identify the space $\mscrS^{\La}_{\!\! g, \scn}$ with $\mscrS^{\La}_{\! 0, \scn}$ for any value of $g\,$, so the perturbative approach can be used without worries. Of course the problem of divergences comes back when taking the limit $\La \to \infty$. Van Hove points out that this is not necessarily true for \emph{every} observable, since there may be quantities whose perturbative estimate remains meaningful in the $\La \to \infty$ limit.

\subsection{Two examples}
\label{subsec: Two examples}

To illustrate this point he considers the interaction energy $B$ between the sources, and the absolute value $a$ of the scalar product between the free and the interacting ground state. Regarding the former, after imposing the cut-of\mbox{}f we can use the usual time-independent perturbation method familiar from Quantum Mechanics (see e.g.\ \cite{Schiff 1968}) to compute the ground state energy to second order in $g$, obtaining
\begin{equation}
E_{^{\{ 0\}, \La}}^{^{(2)}} = - g^{2} \Bigg[ \, \frac{q_{_{1}}^{2} + q_{_{2}}^{2}}{2 V} \sum_{\veck \in \mcalR_{_{\La}}} \! \frac{1}{\o_{k}^{2}} \, + \frac{q_{_{1}} q_{_{2}}}{V} \sum_{\veck \in \mcalR_{_{\La}}} \! \frac{\cos \big[ \veck \cdot (\vecy_{_{1}} - \vecy_{_{2}}) \big] }{\o_{k}^{2}} \, \Bigg] \quad .
\end{equation}
Van Hove identif\mbox{}ies the interaction energy between the sources by ``excluding every transition where one boson is emitted and absorbed by the same source''. Since those transitions would give rise to terms quadratic in $q_{_{1}}$ and in $q_{_{2}}\,$, according to this criterion only the term containing the product $q_{_{1}} q_{_{2}}$ survives. The second order correction $B_{^{\La}}^{^{(2)}}$ to the interaction energy between the sources then takes the form
\begin{equation}
B_{^{\La}}^{^{(2)}} = - \Big( g \, F_{^{\La}}^{^{(2)}} \! \Big)^{\! 2} \quad ,
\end{equation}
where
\begin{equation}
F_{^{\La}}^{^{(2)}} = \Bigg[ \, \frac{q_{_{1}} q_{_{2}}}{V} \sum_{\veck \in \mcalR_{_{\La}}} \! \frac{\cos \big[ \veck \cdot (\vecy_{_{1}} - \vecy_{_{2}}) \big] }{\o_{k}^{2}} \, \Bigg]^{\frac{1}{2}} \quad .
\end{equation}
Reminding the results of section \ref{sec: stationary states} it is clear that $F_{^{\La}}^{^{(2)}} \! $ has a f\mbox{}inite and non-zero limit when $\La \to \infty$, since $\lim_{\La \to \infty} \, g \, F_{^{\La}}^{^{(2)}} = - \sqrt{B}\,$.

On the other hand, indicating with $\Psi_{^{\{0\}}}\! $ the ground state of the free theory ($g = 0$) and with $\Phi_{^{\{0\}}}\! $ the ground state of the interacting one (with the sources at the same positions in the two cases), let us consider $a = \abs{\langle \Phi_{^{\{0\}}} \!\! \mid \!\! \Psi_{^{\{0\}}} \rangle}\,$. Again imposing the cut-of\mbox{}f and using the time-independent perturbation formulas, we obtain to second order
\begin{equation}
a_{^{\La}}^{_{(2)}} = 1 - \Big( g \, G_{^{\La}}^{^{(2)}} \! \Big)^{\! 2} \quad ,
\end{equation}
where
\begin{equation}
G_{^{\La}}^{^{(2)}} = \Bigg[ \, \frac{1}{4 V} \sum_{\veck \in \mcalR_{_{\La}}} \! \frac{1}{\o_{k}^{3}} \, \Big( q_{_{1}}^{2} + q_{_{2}}^{2} + 2 \, q_{_{1}} q_{_{2}} \cos \big[ \veck \cdot (\vecy_{_{1}} - \vecy_{_{2}}) \big] \Big) \, \Bigg]^{\frac{1}{2}} \quad .
\end{equation}
This quantity instead diverges in the limit $\La \to \infty\,$, which brings van Hove to conclude that ``the perturbative approach is then unreliable: it would give a series of the form $1 - g^{2} \cdot \infty + g^{4} \cdot \infty - \ldots$ . The exact value of $a$ is however very simple: $a = 0\,$.''

\section{$S$ matrix and renormalization}
\label{sec: S matrix and renormalization}

In the last section of the article, van Hove (under the suggestion of W.\ Pauli) applies the renormalization procedure to the model studied in the previous sections. The aim is to investigate the reliability of perturbative renormalization by comparing the renormalized result with the exact one. To adhere to the contemporary taste we recast van Hove's analysis in terms of the Wick theorem, while the original was based on the method by Coester and Jauch \cite{Coester 1949}. Apart from this we follow van Hove's line of thought, leaving comments and criticisms to section \ref{sec: Comments}.

\subsection{Interaction picture}

As usual in perturbative QFT, the analysis is done in the interaction picture. Recall that, to pass from the Schr\"{o}dinger to the interaction picture, each operator $\mcal{O}_{_{\! S}}$ is mapped into a (time-dependent) operator $\mcal{O}_{_{\! I}}(t)$ by the formal transformation
\beq \label{Schroedinger to Interaction}
\hat{\mcal{O}}_{_{\! S}} \to \hat{\mcal{O}}_{_{\! I}}(t) = e^{i t \hH_{_{0}}} \,\, \hat{\mcal{O}}_{_{\! S}} \,\, e^{-i t \hH_{_{0}}} \quad ,
\eeq
where $t = 0$ has been chosen as the reference time when the two pictures coincide. It follows then that the f\mbox{}ield operator in the interaction picture reads
\beq \label{quantised phi int}
\hf_{_{\! I}}(t, \vecx) = \frac{1}{\sqrt{2 V}} \, \sum_{\veck \in \mcalR} \, \frac{1}{\sqrt{\o_{k}}} \,\, \bigg[ \, \ha_{\veck} \,\, e^{i \big( \veck \cdot \vecx - \o_{k}^{\phantom{1}} t \big)} + \ha^{\dagger}_{\veck} \,\, e^{i \big( \o_{k}^{\phantom{1}} t - \veck \cdot \vecx \big)} \bigg] \quad ,
\eeq
and as usual it can be decomposed into its positive and negative frequency parts
\begin{align}
\hf_{_{\! I}}^{+}(t, \vecx) &= \frac{1}{\sqrt{2 V}} \, \sum_{\veck \in \mcalR} \, \frac{1}{\sqrt{\o_{k}}} \,\, e^{i \big( \veck \cdot \vecx - \o_{k}^{\phantom{1}} t \big)} \, \ha_{\veck} \quad , \label{fi positive frequency part} \\[2mm]
\hf_{_{\! I}}^{-}(t, \vecx) &= \frac{1}{\sqrt{2 V}} \, \sum_{\veck \in \mcalR} \, \frac{1}{\sqrt{\o_{k}}} \,\, e^{i \big( \o_{k}^{\phantom{1}} t - \veck \cdot \vecx \big)} \, \ha^{\dagger}_{\veck} \label{fi negative frequency part} \quad .
\end{align}

To ensure convergence of the relevant integrals, and thereby to be able to study the evolution from $t = - \infty$ to $t = + \infty$ and not only for f\mbox{}inite times, van Hove introduces an exponential convergence factor. We therefore consider the ``regulated'' interaction Hamiltonian
\beq
\hH_{_{\!I}}^{\a}(t) = e^{- \a \abs{t}} \, \sum_{i = 1}^{2} \, q_{_{i}} \, \hf_{_{I}}(t, \vecy_{_{i}}) \quad ,
\eeq
where $\a$ is a positive real number which is to be sent to zero at the end of the calculation. The physical interpretation of this procedure is that, starting from the asymptotic past, the interaction is switched on progressively, and then it is progressively switched of\mbox{}f going towards the asymptotic future. The $\a \to 0$ limit means that the switch on-of\mbox{}f is done ``inf\mbox{}initely slowly''.

\subsubsection{Perturbative expansion}

Using the Dyson expansion \cite{Dyson 1949}, the unitary operator $U_{^{\!I}}^{\a}(t, - \infty)$ which evolutes the system in the interaction picture from the asymptotic past to the time $t$ can be written in the form
\begin{equation} \label{evolution operator}
\hU_{_{\!I}}^{\a}(t, - \infty) = \sum_{n = 0}^{+ \infty} \, \frac{1}{n!} \, \Big( \frac{g}{i} \Big)^{\! n} \! \int_{- \infty}^{t} d t_{_{1}} \ldots \int_{- \infty}^{t} d t_{_{n}} \, T \Big\{ \hH_{_{\!I}}^{\a}(t_{_{1}}) \cdots \hH_{_{\!I}}^{\a}(t_{_{n}}) \Big\} \quad ,
\end{equation}
where $T \{\ldots\}$ indicates the time-ordered product where the factors are ordered from the left to the right according to decreasing time.

Due to the simple structure of the interaction, which is is linear in the f\mbox{}ield, the interaction Hamiltonian itself decomposes into a positive and a negative frequency part
\begin{equation}
\hH_{_{\!I}}^{\a}(t) = \hH_{\!+}^{\a}(t) + \hH_{\!-}^{\a}(t) \qquad , \qquad \hH_{\!\pm}^{\a}(t) = e^{- \a \abs{t}} \, \sum_{i = 1}^{2} \, q_{_{i}} \, \hf_{_{\! I}}^{\pm}(t, \vecy_{_{i}}) \quad .
\end{equation}
The Wick theorem can then be applied directly to the interaction Hamiltonians, to pass from the time-ordered product to the normal-ordered product $N \{ \hH_{_{\!I}}^{\a}(t_{_{1}}) \cdots \hH_{_{\!I}}^{\a}(t_{_{n}}) \}$ where the negative frequency parts are placed to the left with respect to the positive frequency ones. Moreover, as a consequence of the linearity of the interaction there are no loop integrations/diagrams, and, whenever performing contractions in one time-ordered product, the time integrals of the contractions factor out. As we show in appendix \ref{App: Factorization}, the contribution from the contractions completely factors out from the sum over $n\,$, and the whole time evolution operator can be compactly written as
\begin{multline} \label{evolution operator factorized}
\hU_{_{\!I}}^{\a}(t, - \infty) = \exp \Bigg( \! - \frac{g^{2}}{2} \int_{- \infty}^{t} \!\! d \t_{_{1}} \!\! \int_{- \infty}^{t} \!\! d \t_{_{2}} \, \acontraction[1ex]{}{\hH_{_{\!I}}^{\a}}{(\t_{_{1}})}{\hH_{_{\!I}}^{\a}} \hH_{_{\!I}}^{\a}(\t_{_{1}}) \, \hH_{_{\!I}}^{\a}(\t_{_{2}}) \Bigg) \, \cdot \\
\cdot \Bigg( \sum_{n = 0}^{+ \infty} \, \frac{1}{n!} \, \Big( \frac{g}{i} \Big)^{\! n} \! \int_{- \infty}^{t} \!\! d t_{_{1}} \, \ldots \int_{- \infty}^{t} \!\! d t_{_{n}} \, N \Big\{ \hH_{_{\!I}}^{\a}(t_{_{1}}) \cdots \hH_{_{\!I}}^{\a}(t_{_{n}}) \Big\} \Bigg) \quad ,
\end{multline}
where
\begin{equation*}
\acontraction[1ex]{}{\hH_{_{\!I}}^{\a}}{(t_{_{1}})}{\hH_{_{\!I}}^{\a}} \hH_{_{\!I}}^{\a}(t_{_{1}}) \, \hH_{_{\!I}}^{\a}(t_{_{2}}) = \frac{e^{-\a ( \abs{t_{_{1}}} + \abs{t_{_{2}}} ) }}{2V} \sum_{i = 1}^{2} \sum_{j = 1}^{2} q_{_{i}} q_{_{j}} \sum_{\veck \in \mcalR} \, e^{i \veck \cdot (\vecy_{i} - \vecy_{j})} \, \frac{e^{- i \, \o_{k}^{\phantom{1}} \abs{t_{_{1}} - t_{_{2}}}}}{\o_{k}^{\phantom{1}}} \quad .
\end{equation*}

\subsection{Renormalization}

Let us consider now the $S$ matrix for our system, that is $\hat{S}^{\a} = \hU_{^{\!I}}^{\a}(+ \infty , - \infty)\,$. It is not dif\mbox{}f\mbox{}icult to see that the double integral of the contraction becomes purely imaginary in the limit $\a \to 0\,$, and that
\begin{equation} \label{infinite phase}
\lim_{\a \to 0} \, - \frac{g^{2}}{2} \int_{- \infty}^{\infty} \!\! d \t_{_{1}} \!\! \int_{- \infty}^{\infty} \!\! d \t_{_{2}} \, \acontraction[1ex]{}{\hH_{_{\!I}}^{\a}}{(\t_{_{1}})}{\hH_{_{\!I}}^{\a}} \hH_{_{\!I}}^{\a}(\t_{_{1}}) \, \hH_{_{\!I}}^{\a}(\t_{_{2}}) \sim - i \! \int_{- \infty}^{\infty} \! \big( B + C \big) \,\, d t \quad , 
\end{equation}
where $B$ and $C$ are those of equations (\ref{Yukawa int energy}) and (\ref{Yukawa self-energy}). In other words, when $\a \to 0$ the fully contracted terms give rise to an inf\mbox{}inite phase in the S matrix, equal to the time integral of the energy of the vacuum. On the other hand, the term in $\hat{S}^{\a}$ with the normal ordered f\mbox{}ields has f\mbox{}inite matrix elements between any two (f\mbox{}inite norm) states of the free theory, even in the limit $\a \to 0\,$. Van Hove then declares that the only necessary renormalization is to redef\mbox{}ine the interaction Hamiltonian in such a way to absorb the inf\mbox{}inite energy $B + C\,$. In other words, he maintains that in this case the renormalization reduces to a rigid displacement of the energy of an inf\mbox{}inite amount. In the process the phase factor disappears altogether, and he obtains the renormalized $S$ matrix
\begin{equation} \label{renormalized S matrix}
\hS_{\a}^{\, \textup{ren}} = \sum_{n = 0}^{+ \infty} \, \frac{1}{n!} \, \Big( \frac{g}{i} \Big)^{\! n} \! \int_{- \infty}^{\infty} \!\! d t_{_{1}} \, \ldots \int_{- \infty}^{\infty} \!\! d t_{_{n}} \, N \Big\{ \hH_{_{\!I}}^{\a}(t_{_{1}}) \cdots \hH_{_{\!I}}^{\a}(t_{_{n}}) \Big\} \quad ,
\end{equation}
and the renormalized (f\mbox{}inite time) evolution operator
\begin{equation} \label{renormalized evolution operator}
\hU_{\a}^{\, \textup{ren}}(t, - \infty) = \sum_{n = 0}^{+ \infty} \, \frac{1}{n!} \, \Big( \frac{g}{i} \Big)^{\! n} \! \int_{- \infty}^{t} d t_{_{1}} \ldots \int_{- \infty}^{t} d t_{_{n}} \, N \Big\{ \hH_{_{\!I}}^{\a}(t_{_{1}}) \cdots \hH_{_{\!I}}^{\a}(t_{_{n}}) \Big\} \quad .
\end{equation}

\subsubsection{Renormalized \emph{vs} exact result}

Consider now two states of the free theory, the ground state $\mid \! 0 \, \rangle$ and the one-particle state $\mid \! \veck \, \rangle$ of momentum $\veck \neq \vec{0} \,$, and evaluate the matrix element of $\hU_{\a}^{\, \textup{ren}}(t, - \infty)$ between these states. Only the linear term in the expansion (\ref{renormalized evolution operator}) contributes to this process, and we get \footnote{Recall that $\th(t) = 1$ for $t > 0$ and $= 0$ for $t < 0\,$, while $\text{sgn}(t) = \pm 1$ for $t \gtrless 0\,$. We use the convention $\th(0) = 1/2$ and $\text{sgn}(0) = 0\,$.}
\begin{multline}
\langle \, \veck \mid \hU_{\a}^{\, \textup{ren}}(t, - \infty) \mid 0 \, \rangle = \\
= - \frac{i}{\sqrt{2 V}} \,\, \sum_{i = 1}^{2} \, g_{_{i}} \, \frac{e^{- i \, \veck \cdot \vecy_{_{i}}}}{\sqrt{\o_{k}}} \, \frac{2 \a \, \th(t) - \big( i \, \o_{_{k}} + \a \, \sgn(t) \big) \, e^{- \a \abs{t}} \, e^{i \, \o_{k}^{\phantom{1}} t} }{\a^{2} + \o_{^{k}}^{2}} \quad .
\end{multline}
Taking the limit $\a \to 0\,$, it is then easy to see that the matrix element is non-vanishing for $t$ f\mbox{}inite:
\begin{equation} \label{finite t transition amplitude}
\lim_{\a \to 0} \, \langle \, \veck \mid \hU_{\a}^{\, \textup{ren}}(t, - \infty) \mid 0 \, \rangle = - \frac{e^{i \, \o_{k}^{\phantom{1}} t}}{\sqrt{2 V \o_{k}^{3}}} \,\, \sum_{i = 1}^{2} \, g_{_{i}} \, e^{- i \, \veck \cdot \vecy_{_{i}}} \neq 0 \quad ,
\end{equation}
while it vanishes for $t = + \infty\,$:
\begin{equation}
\lim_{\a \to 0} \, \langle \, \veck \mid \hS_{\a}^{\, \textup{ren}} \mid 0 \, \rangle = 0 \quad .
\end{equation}
More in general, all the of\mbox{}f-diagonal elements of $\hS_{\a}^{\, \textup{ren}}$ vanish in the $\a \to 0$ limit.

In comparing the predictions of the renormalized perturbative expansion with the exact result of section \ref{sec: stationary states}, recall that in the latter case the interaction causes the energy levels to shift by the (inf\mbox{}inite) amount $B + C$  with respect to the free case. Apart from this, the energy spectrum is unaf\mbox{}fected by the interaction, as equation (\ref{spectrum}) shows (note that $C$ has been neglected there). Moreover, van Hove points out that the interaction changes the form of the modes (equations \eqref{varphi} and \eqref{eigenfunctions}) near the sources (or better, in the non-asymptotic region), but ``does not introduce a coupling between them''. These features are compatible with the picture given by the renormalized perturbative expansion, where a rigid displacement of the energy of the amount $B + C$ appears and where the $S$ matrix is the identity. Bear in mind that taking the limit $\a \to 0$ is not merely a technical detail, since in presence of the convergence factor $e^{- \a \abs{t}}$ the conf\mbox{}igurations of section \ref{sec: stationary states} are not anymore exact solutions of the equations of motion, so we cannot expect the exact results to be recovered away from this limit.

On the other hand, van Hove states that ``it is more dif\mbox{}f\mbox{}icult to interpret the meaning of the non-vanishing of\mbox{}f-diagonal elements of $\lim_{\a \to 0} U_{\a}^{\, \textup{ren}}(t, - \infty)$ for f\mbox{}inite $t$''. He takes as an example the expression (\ref{finite t transition amplitude}) which seems to imply that, preparing the system in the ground state in the inf\mbox{}inite past, there is a non-zero probability of detecting an excited mode if we perform a measure at the f\mbox{}inite time $t\,$. This in principle could be ascribed as a spurious ef\mbox{}fect of switching on and of\mbox{}f the interaction, however in the limit $\a \to 0$ the switching on and of\mbox{}f is so slow that we expect the system to respond adiabatically. The perturbative prediction of excitations from the ground state brings van Hove to say that ``nothing in the exact solution suggests the possibility of such a phenomenon, and, quite generally, there are reasons to doubt of the predictions based on the matrix elements of $\lim_{\a \to 0} U_{\a}^{\, \textup{ren}}(t, - \infty)$ for f\mbox{}inite $t\,$, as obtained via the renormalization procedure''. He reinforces this point stating that ``it is important to note how incomplete is the description of the f\mbox{}ield provided by this procedure''.

\section{Comments}
\label{sec: Comments}

The reading of van Hove's article inspires several interesting remarks and some criticism. In this section, we take advantage of the benef\mbox{}its of hindsight and comment on the points of the article which are, in our opinion, the most interesting, placing them inside the contemporary context.

\subsection{Quantization methodology}

We start by commenting on the quantization methodology. In introductory QFT textbooks (such as, for example, \cite{Peskin Schroeder}), the canonical quantization of a neutral scalar f\mbox{}ield is usually presented as follows. The classical f\mbox{}ield is decomposed into Fourier components, in such a way that the positive/negative frequency decomposition is already apparent by suitably naming half of the coef\mbox{}f\mbox{}icients, say, by $a\,$, and the other half, say, by $a^{\ast}$. At this point the quantization is implemented by promoting the Fourier coef\mbox{}f\mbox{}icients to operators ($a \to \ha$ and $a^{\ast} \to \ha^{\dag}$) which are to satisfy the commutation relations (\ref{a adg commutation relations}), and contextually the space of state vectors is constructed at the abstract level by means of these same operators, by assuming the existence of a ``vacuum state'' (or ``no particle'' state) annihilated by all the $\ha\,$.

\subsubsection{Representations and ambient space}

What is actually being done is the construction of a Fock representation of the CCR.\footnote{Recall that CCR = Canonical Commutation Relations and UIR = Unitarily Inequivalent Representation(s).} Although there are mathematical subtleties involved in the construction as sketched above, it can be made mathematically precise \cite{Cook 1953}. What is usually not stated clearly is that in the process we are choosing one over an inf\mbox{}inite number of UIR of the CCR, nor it is discussed why we are neglecting all the others (this matters are usually discussed only in more advanced, and mathematically oriented, books on QFT, and by the community of philosophers of physics). It is then implicitly assumed that the quantum dynamics of the system happens always inside that representation, independently of the form of the interaction.

Van Hove's analysis departs from the path sketched above when he introduces the family of operators $\hq$ and $\hp$ in (\ref{qk pk}), which obey the CCR (\ref{q p CR}). Although at f\mbox{}irst this is just a redef\mbox{}inition, it paves the way for the representation choice of section \ref{subsubsec: Representation}, where the operators $\hq$ and $\hp$ are represented as dif\mbox{}ferential operators on a space of functionals (relations (\ref{functional}) and (\ref{differential})).\footnote{It is interesting to speculate as the passage from $\ha$ and $\ha^{\dg}$ to $\hq$ and $\hp$ and then to the operators (\ref{qk pk}) can be seen as a ``de-algebrization''. Recall that, in the quantum theory of the harmonic oscillator, the Hamiltonian and the eigenvalue problem are usually introduced in the Schr\"{o}dinger formulation, therefore as a dif\mbox{}ferential problem. It is then recognized that, introducing the creation and annihilation operators, the problem can be most easily dealt with by purely algebraic means (with a bit of caution on the domain of the operators). In the article, van Hove does exactly the opposite, performing the quantization in an abstract way and then transforming the algebraic problem into a dif\mbox{}ferential one. The analogy cannot be pushed too far, however, since the operators $\hq$ do not correspond to the position coordinates of a physical system, but are coordinates in a space of functions.} 
The crucial point is that the functional space on which these operators act is not a priori identif\mbox{}ied with the space of state vectors of the system. In other words, the space of state vectors is not imposed by hand rigidly when quantizing (tailored on the non-interacting case), but the Hamiltonian is given free rein to individuate it as a subspace of a wider space. This is the key passage which allows to discover that Hamiltonians relative to dif\mbox{}ferent choices of the parameters individuate dif\mbox{}ferent spaces of state vectors, in fact one orthogonal to the other as the article goes on to show.

\subsubsection{Separable \textit{vs} non-separable Hilbert spaces}

From a wider perspective, the key point is that van Hove chooses to represent the CCR on a \emph{non-separable} Hilbert space, while usually (as in the Fock case) one looks for representations on a separable Hilbert space. From a physical point of view, the preference towards separable Hilbert spaces is arguably connected to the idea that the ontology underlying quantum f\mbox{}ields is one of particles, since in that case one expects that the states which correspond to aggregations of particles (which constitute a countable set) span the whole space of state vectors. The non-separable Hilbert space considered by van Hove is more specif\mbox{}ically obtained as the inf\mbox{}inite direct product of separable Hilbert spaces. The crucial point in this regard is that, as von Neumann showed \cite{von Neumann Comp Math 1939}, such a space decomposes into an uncountable direct sum of separable Hilbert spaces, each of which is therefore isomorphic to the usual Fock space. It is this property which permits to have as many spaces of state vectors as the possible choices for the parameters of the model, each of them orthogonal to the others, and nonetheless being separable Hilbert spaces themselves.

It is worthwhile to recall how both Wightman (\cite{Streater Wightman}, pages 86--87) and Wald (\cite{Wald book}, page 33), while explicitly citing von Neumann's inf\mbox{}inite direct product construction, argument that a separable Hilbert space should provide a suf\mbox{}f\mbox{}icient structure to formulate a QFT. A posteriori, van Hove's analysis suggests that a separable Hilbert space is indeed the natural framework for a QFT, if by this we mean a QFT Hamiltonian with a f\mbox{}ixed choice of its parameters. However, if we want to be able to compare the predictions of a class of Hamiltonians, e.g.\ the one obtained by varying the parameters of a given Hamiltonian as is done in perturbation theory, then a single separable Hilbert space is not an adequate choice anymore. It is more convenient to have all the class coexist in a single space, which then need to be much wider and therefore non-separable.

\subsection{Historical role}

We now turn to a comment on the historical importance of the article, more precisely to the role it had in spreading into the community of physicists the awareness of the importance of UIR of the CCR.

\subsubsection{Unitarily inequivalent representations of the CCR}

The problem of the existence and the relations between dif\mbox{}ferent representations of the CCR appeared rather early in the history of QM, due to the appearance of two competing formulations of the theory (Heisenberg, Born and Jordan's and Schr\"{o}dinger's). As we mentioned in section \ref{sec: CCR interaction picture}, for quantum systems with a f\mbox{}inite number of degrees of freedom there are several mathematical results which satisfactorily settle the matter. The best known of these results, to which we referred collectively as unitary equivalence theorems, is the Stone-von Neumann theorem.

Although these theorems fail for systems with inf\mbox{}inite degrees of freedom, like f\mbox{}ields, the role of the representations of the CCR in QFT did not attract a lot of attention at f\mbox{}irst. Even the very notion of the existence of representations dif\mbox{}ferent from the Fock's one was slow to dif\mbox{}fuse into the community of physicists. Indeed, although the existence of UIR of the CCR where known to some at least before World War II \cite{von Neumann Comp Math 1939} (to von Neumann, at any rate), the community of physicists started to become more widely aware of their relevance only in the 1950s.

It is in this context that the historical importance of van Hove's article is more pronounced. Together with the work of Friedrichs \cite{Friedrichs 1952}, the article is credited both by Wightman and Schweber \cite{Wightman Schweber 1955} and Haag \cite{Haag 1955} as the main sources which made the existence of the UIR of the CCR enter into the consciousness of physicists. Wightman and Schweber explicitly recognized being strongly inf\mbox{}luenced by van Hove's article saying that (\cite{Wightman Schweber 1955}, page 824) ``our illustrations of these points are based on the fundamental work of van Hove, and our discussion may be regarded as an alternative explication of his results''. The importance of Wightman's and Haag's contribution to the development of QFT, including the cited works, witnesses the key role that van Hove's article had in opening a new avenue of investigation in the quantum theory of f\mbox{}ields.

\subsubsection{Haag's theorem}

A word is in order about the relationship between van Hove's article and Haag's theorem. With the latter name it is customary to refer to a set of results, starting with Haag's own one in \cite{Haag 1955} and comprising generalizations and reformulations thereof (arguably, the best known being that of Hall and Wightman \cite{Hall Wightman 1957}). See \cite{Lupher 2005} for an historical survey and \cite{Earman Fraser 2006} for a philosophical discussion.

In its original form, the theorem goes as follows. Consider $\{\hps(\vecx), \hpi(\vecx)\}$ and $\{\hvf(\vecx), \hr(\vecx)\}$ two separable Hilbert space representations of the CCR for a scalar f\mbox{}ield
\begin{align*}
\Big[ \, \hps(\vecx) \, , \hps(\vecy \, ) \Big] &= \Big[ \, \hpi(\vecx) \, , \hpi(\vecy \, ) \Big] = 0 \quad , & \Big[ \, \hps(\vecx) \, , \hpi(\vecy \, ) \Big] &= i \, \d(\vecx - \vecy \, ) \quad , \\[2mm]
\Big[ \, \hvf(\vecx) \, , \hvf(\vecy \, ) \Big] &= \Big[ \, \hr(\vecx) \, , \hr(\vecy \, ) \Big] = 0 \quad , & \Big[ \, \hvf(\vecx) \, , \hr(\vecy \, ) \Big] &= i \, \d(\vecx - \vecy \, ) \quad ,
\end{align*}
where $\hpi$ and $\hr$ are the canonical momenta associated respectively to $\hps$ and $\hvf\,$. Suppose they are unitarily equivalent, so there exists a unitary operator $\hat{R}$ such that
\begin{align} \label{unitary equivalence}
\hps(\vecx) &= \hat{R} \,\, \hvf(\vecx) \,\, \hat{R}^{\dag} \qquad , & \hpi(\vecx) &= \hat{R} \,\, \hr(\vecx) \,\, \hat{R}^{\dag} \quad ,
\end{align}
for every $\vecx \in \mbbR^{3}$. If the system is translationally invariant, then $\hat{R}$ commutes with the generators of the translations, that is with the momentum operators
\begin{equation}
\Big[ \, \hat{R} \, , \hat{P}_{i} \, \Big] = 0 \qquad , \qquad i = 1, 2, 3 \quad .
\end{equation}
Assuming translational invariance, suppose moreover that the operators $\hat{P}_{i}$ have continuous spectrum, with the only exception of the non-degenerate discrete eigenvalue $(0,0,0)$. Calling $\Phi_{_{0}}$ a normalized eigenvector of $\hat{P}_{i}$ with respect to the eigenvalue $(0,0,0)$, then it follows that $\hat{R} \, \Phi_{_{0}} = \Phi_{_{0}}$ (up to phase).

The deep implication of this result becomes clear if we suppose that both representations $\{\hps(\vecx), \hpi(\vecx)\}$ and $\{\hvf(\vecx), \hr(\vecx)\}$ (which may correspond to a non-interacting theory and a self-interacting one) have a vacuum state which is translationally invariant. Then necessarily the two theories share the \emph{same} vacuum state. In other words, borrowing the terminology of renormalized QFT, the interaction cannot ``polarize'' the free vacuum state. This conclusion contradicts almost everything we know from the perturbative analysis of realistic QFT models, whose predictions are extremely well conf\mbox{}irmed by experiments. More ref\mbox{}ined versions of the theorem generalize the result to the ef\mbox{}fect that not only a non-interacting and a self-interacting theory share the same vacuum state, but they are actually the \emph{same theory}: provided the hypotheses hold, the only theory unitarily equivalent to a free theory is a free theory. This in particular undermines the existence of the interaction picture in QFT, whenever the theorem applies. Among the assumptions on which the theorem and its generalizations rest, the easiest to circumvent is the translational invariance of the theory. To escape its conclusions it is in fact suf\mbox{}f\mbox{}icient to put the system in a box, or to make it interact with external sources having a non-trivial prof\mbox{}ile.\footnote{Incidentally, this is why Haag's theorem does \emph{not} undermine the validity of the calculations in renormalized perturbative QFT. As Duncan (\cite{Duncan 2012}, section 10.5) explains, the introduction of a regularizing procedure usually evades the hypotheses of the theorem.\label{Duncan footnote}}

It is clear from what said above that van Hove's model does not satisfy the hypotheses of the theorem, since translational invariance is broken both by the f\mbox{}ield interacting with sources which lie at f\mbox{}ixed positions, and by the system living in a box. Interestingly we may say that, regarding the relevance of UIR of the CCR to QFT, the relation between van Hove's model and Haag's theorem is one of complementarity. The latter gives quite general and natural conditions, although easily circumvented, under which describing interaction in QFT calls for the use of UIR of the CCR. The former instead is a very specif\mbox{}ic model, which is artif\mbox{}icial in several ways, but points to the same conclusion about interaction and representations, even beyond the validity of the theorem. In particular, van Hove's model makes it clear that the case for considering UIR of the CCR in QFT is stronger than Haag's theorem makes it.

\subsection{The perturbative approach in QFT}

Arguably, the farthest-reaching point of van Hove's discussion is given by his questioning the validity of the perturbative approach in QFT, and the connection he suggests between the phenomenon of orthogonality of the spaces of state vectors and the appearance of divergences in perturbative QFT. As these puzzled the community of physicists for a long time, even leading some to propose that the formalism should be abandoned altogether, a comment on this matter is in order.

Divergences are commonly divided into infrared and ultraviolet ones, according to whether they are due to modes with momenta tending to zero or to inf\mbox{}inity. It is known that the inf\mbox{}inite volume limit in QFT naturally leads to UIR of the CCR \cite{Friedrichs 1952} (and also in ordinary QM, when the thermodynamic limit is considered). However, since in van Hove's model the system lives in a box of f\mbox{}inite volume, we may expect his comments to be directed mainly towards ultraviolet divergences, although he does not say so explicitly. We assume this to be the case.

\subsubsection{Some general comments}

We start with short comments on two often encountered assertions concerning QFT and divergences.

The f\mbox{}irst one is the assertion that divergences in realistic QFT are due to (and inescapable consequence of) the number of degrees of freedom being inf\mbox{}inite.\footnote{The adjective ``realistic'' is not superf\mbox{}luous here, since it is known that f\mbox{}inite QFT exist in lower dimensionality, for example in $1 + 1$ spacetime dimensions \cite{Glimm Jaffe I}.} Van Hove's discussion emphasizes that the foremost consequence of changing the number the degrees of freedom from f\mbox{}inite to inf\mbox{}inite, is a change in the structure of the spaces of state vectors. Moreover, he points out that spurious divergences are very likely to be introduced if this structure is not adequately taken into account when using the perturbative approach. This leaves open the possibility for a realistic quantum theory with an inf\mbox{}inite number of degrees of freedom not being divergent at all, in case all the perturbative divergences turn out to be spurious (in the sense of the previous sentence). This is of course a very long shot. But, at the very least van Hove's analysis implies that, to establish whether the quantum description of a system with an inf\mbox{}inite number of degrees of freedom is divergent or not, it is necessary to properly take into account the structure of the spaces of state vectors.

The second one is the assertion that ultraviolet divergences are to be expected because the theory won't describe nature up to inf\mbox{}inite energy. Although it is very sensible that QFT (and specif\mbox{}ically the Standard Model) may not describe nature correctly up to arbitrary high energies, we observe that its eventual high energy inadequacy does not need to manifest itself through divergences. It may also manifest as a f\mbox{}inite mismatch between the QFT predictions and the experimental data or as internal inconsistencies of the formalism other than divergences. What we are saying is that, although we do expect new physics to turn up, it is incorrect to propose a \emph{strict} logical equivalence between divergences and high frequency inadequacy of the theory. And, that it is wrong to philosophically justify the presence of divergences as unavoidable on the basis of the fact that we don't expect our description to hold up to inf\mbox{}inite energy.

Having said that, a possible link between ultraviolet divergences and the phenomenon of orthogonality of the spaces of state vectors directly calls into question the renormalization procedure, since the latter was developed exactly to tame ultraviolet divergences in the perturbative expansion. The main question which stems from this suggestion is what would the consequences for renormalization be if we formulated QFT in a non-separable Hilbert space. Simplifying: would there still be ultraviolet divergences? If yes, which aspects of renormalization would survive? Without aiming to provide an answer, some (admittedly speculative) considerations can be elaborated. In the section below we push the speculation as far as it can go, supposing that ultraviolet divergences can be wholly explained by the phenomenon of orthogonality of the spaces of state vectors, for the sake of checking whether we run into plain contradiction.

\subsubsection{Renormalization and orthogonality}

The fact that divergences in non-linear QFT are ubiquitous may testify in favor of a structural problem at the their basis, such as the choice of having the free and interaction Hamiltonians live in a space which is not wide enough. It is interesting to point out that, if this were the case, the divergences would likely not be associated to a specif\mbox{}ic energy domain, but to the global choice of the space. This immediately clashes with the common lore in the old approach to renormalization, where the divergences were associated to the high energy behavior of the theory (``failure above certain frequencies'') \cite{Cao Schweber Synthese 1993}.

It is important to remark that, even if the divergences were of ``global'' origin we could make them \emph{appear} as a high energy phenomenon. It would be enough to impose a high-energy cut-of\mbox{}f and take advantage of the Stone-von Neumann theorem on the remaining (f\mbox{}inite) degrees of freedom (this is in fact what is usually done in QFT). That is, we would map the states and the Hamiltonian of the (cut-of\mbox{}f) interacting theory on the space of state vectors of the free theory, therefore \emph{away} from the true interacting space of state vectors. The divergence encountered when sending the cut-of\mbox{}f to inf\mbox{}inity would then be artif\mbox{}icially created by this procedure, and appear as a high frequency phenomenon only because we implemented the recovery of the full theory as a high frequency limit. It is interesting in this respect that, in von Neumann's paper \cite{von Neumann Comp Math 1939} on inf\mbox{}inite direct products of vector spaces, convergence is not def\mbox{}ined as a limit when some number goes to inf\mbox{}inity. On the contrary, it is def\mbox{}ined without choosing a preferred ordering in the index set which label the spaces (in our case the index set would be the integer numbers which label the modes of the f\mbox{}ield, which are tightly linked to their frequency). 

The modern approach to renormalization, on the other hand, takes a dif\mbox{}ferent point of view in that the focus is not exclusively on the high energy behavior and on ways of circumventing the divergences. In fact, borrowing David Gross' words \cite{Gross 1985}, ``renormalization is an expression of the variation of the structure of physical interactions with changes in the scale of the phenomena being probed'', and borrowing Michael Dine's \cite{Dine 2015} ``renormalization is the statement that the parameters of a theory vary with length or energy scale''. All the same, it may seem that the cited dependence on the scale being probed is again at odds with the possibility that the divergences be due to the orthogonality between the space of state vectors, since the latter phenomenon being of global nature is not expected to produce an energy dependence.

This conclusion would be, however, premature. It is fair to expect that the features of the renormalized perturbative expansion, if at all, would be reproduced in the context of an appropriate perturbative treatment in the non-separable Hilbert space. Such a perturbative treatment, if it existed, would clearly have to take into account that also the very space where the Hamiltonian lives is perturbed. It seems not impossible that the dependence of the parameters on the energy scale being probed would result from an analysis of the renormalization group type of the perturbative treatment in the non-separable Hilbert space.

\subsection{The comparison with renormalization}

We conclude the present section by discussing the comparison between the exact results and those of renormalized perturbative QFT, exposed in section \ref{sec: S matrix and renormalization}, which in our opinion is the less convincing. Before exposing our main criticism, let us start with a methodological comment.

The analysis in section \ref{sec: S matrix and renormalization} is based on the interaction picture, which assumes that the vector states of the interacting theory can be obtained from the states of the free theory by a unitary transformation acting on the space of state vectors of the latter. This evidently clashes with the results of section \ref{sec: space of the state vectors}, where it is proved that such a transformation cannot exist since the two spaces of state vectors (free and interacting) are not only dif\mbox{}ferent but even orthogonal. Note that in section \ref{sec: S matrix and renormalization} we are \emph{not} imposing a cut-of\mbox{}f, or another regularization procedure, when dealing with the perturbative expansion, so the observation in the footnote on page \pageref{Duncan footnote} does not apply. Therefore, from the methodological point of view the approach of van Hove in section \ref{sec: S matrix and renormalization} is to buy the full package of perturbative QFT without questioning it (the questioning has been already done in section \ref{sec: QFT and perturbative}, in fact), and just apply it to compare its predictions with those of the exact analysis.

\subsubsection{Criticism}

Now for the criticism. A f\mbox{}irst point is that van Hove seems to implement renormalization in a dif\mbox{}ferent way compared to how it is done nowadays. In general, the exponential ``convergence factor'' is not used merely as a way to make integral converge, but as a way to generate the eigenstates of the interacting theory from those of the free theory by means of a suitable time evolution of the system (see e.g.\ \cite{Fetter Walecka 1973}). In this context, taking the limit $\a \to 0$ corresponds to generating the interacting states adiabatically. The inf\mbox{}inite phase factor is an unavoidable consequence of this procedure \cite{Nenciu Rasche 1989}, not a pathology, and in general it is not renormalized away, but taken care of directly in the def\mbox{}inition of the adiabatically-generated interacting states (the relevant result being the Gell-Mann and Low formula \cite{Gell-Mann Low 1951}). Note in particular that the divergence of the phase factor is not due to the energy $B + C$ being divergent, since (as equation (\ref{infinite phase}) shows) the phase is the integral of $B + C$ over time (from $- \infty$ to $+ \infty$) which would diverge also if $B + C$ were f\mbox{}inite and non-zero. We reiterate that loop divergences are not present in the model considered here, as a consequence of the linearity of the interaction.

A second point regards van Hove's conclusion that the renormalized perturbative expansion does not correctly reproduce the exact result for $t$ f\mbox{}inite, exemplif\mbox{}ied by the fact that $\lim_{\a \to 0} U_{\a}^{\, \textup{ren}}(t, - \infty)$ has non vanishing of\mbox{}f-diagonal elements. Van Hove seems to expect that, if in the perturbative treatment the system were prepared in an eigenstate of the free Hamiltonian, then it should stay there for all times. This may be motivated by the idea that the eingenstates of the free Hamiltonian in the perturbative analysis somehow  ``correspond'' to the exact energy eigenstates in the exact analysis. We speculate that such an intuitive idea is what lies behind the assertion that ``the interaction changes the form of the modes, but does not introduce a coupling between them''. From this perspective, in the perturbative treatment the system should remain in an eigenstate of the free Hamiltonian because this is what happens for the ``corresponding'' states in the exact solution.

Anyway, in quantum physics the only relevant quantities are the expectation values of the observables, so the comparison between the exact and the perturbative results should be done on that level. Indeed from that point of view van Hove's conclusion seems questionable. Consider for example the momentum observable, and recall that the exact stationary states are not eigenfunctions of the momentum (as van Hove himself points out), although asymptotically they can be approximated by plane waves. This situation is compatible with the perturbative result (\ref{finite t transition amplitude}), which says that a system prepared at $t = - \infty$ in the (non-interacting) ground state evolves into a \emph{superposition} of momentum eigenstates when the interaction with the sources makes itself felt. On the contrary, if the perturbative analysis predicted that the system would remain in the (non-interacting) ground state for all times, the expectation of the momentum would be constantly zero, in contradiction with the exact result.

It may be suggested that an observable dif\mbox{}ference between the exact and perturbative results lies in the fact that the latter predicts ``particle production'' for $t$ f\mbox{}inite, while the former does not. This argument is not straightforward however, since to make it sound one should couple the f\mbox{}ield with a ``particle detector'', and show that the detector ``clicks'' in the perturbative case and does not in the exact case. Since the exact solutions are not momentum eigenstates, it is not obvious that the detector would not click in the exact case. Such a discussion is missing in the article, so we believe that the analysis of section \ref{sec: S matrix and renormalization} is not suf\mbox{}f\mbox{}icient to support the claim that the exact and perturbative analyses give dif\mbox{}ferent results for f\mbox{}inite time.

\section{Conclusions}
\label{sec: Conclusions}

In \cite{van Hove 1952} L\'{e}on Van Hove studied the model of a neutral, massive and relativistic scalar f\mbox{}ield linearly coupled to point-like sources which lie at f\mbox{}ixed positions. This (admittedly unrealistic) model was chosen for his formal properties of admitting exact solutions, with the purpose of clarifying the origin of the formal problems exhibited by Quantum Field Theory. More specif\mbox{}ically, he tried to understand why the perturbative approach inevitably leads to divergences, and until which point the renormalized perturbative expansion provides a satisfactory approximation of the exact solution.

We are of the opinion that van Hove's analysis, despite the evolution that QFT underwent in the meantime, deserves to be known outside the community of French-speaking mathematical physicists. Not only because of its historical importance, but also because some of the ideas contained in it are still relevant. Moreover, the article has a didactic relevance since it exposes in a simplif\mbox{}ied context a procedure of f\mbox{}ield quantization which is dif\mbox{}ferent from the one usually taught in curricular courses. This was the goal which motivated us to prepare the present exposition, expanding the analysis of \cite{van Hove 1952} with derivations and comments.

Summing up his results, he wrote that ``regarding the f\mbox{}irst problem, the reason lies in the fact that the stationary states of the f\mbox{}ield interacting with the sources are not linear combinations of the stationary states of the free f\mbox{}ield. By virtue of the results of a previous work, this fact extends to the more general case of two interacting quantum f\mbox{}ields, under the only hypothesis that stationary states indeed exist in presence of the interaction.'' Concerning the second point, he says that ``despite they [the exact solution and the results obtained using the renormalization procedure] are in agreement for the $S$ matrix, they disagree with respect to the unitary evolution matrix relative to the time interval $(- \infty, t)$, with $t$ f\mbox{}inite. Furthermore, the renormalization procedure provides an incomplete description of the exact solution.''

With the benef\mbox{}it of hindsight, our opinion about van Hove's conclusions is the following. The f\mbox{}irst point is connected to the existence of Unitarily Inequivalent Representations of the Canonical Commutation Relations in QFT, and it remains a relevant topic. On the other hand, his analysis on the comparison between the exact result and that provided by the renormalized perturbative expansion is, to our judgment, inconclusive.

An interesting question is how these conclusions would change if a less unrealistic model were considered, such as one with sources of f\mbox{}inite but non-zero radius or with a charge prof\mbox{}ile similar to the ground state of the hydrogen atom. The key point is clearly whether the phenomenon of orthogonality of the spaces of state vectors would still be present. A related question is whether it is possible to f\mbox{}ind models where the constant $C$ def\mbox{}ined in (\ref{Yukawa self-energy}), which in the article diverges and is discarded without comments, is f\mbox{}inite (so the new model is not singular) and yet the spaces of state vectors are orthogonal. An investigation on these points is under way, and will be published elsewhere.

\appendix

\section{Proof of the orthogonality results}
\label{App: Orthogonality}

\subsection{First orthogonality result}

Let us suppose that $\G \neq \barG\,$, and evaluate the scalar product
\beq \label{scalar product nm}
\Big\langle \bar{\Phi}_{\{ \barn \}} \, \Big\vert \, \Phi_{\{ n \}} \Big\rangle = \prod_{\veck \in \mcalR} \Big\langle \bar{\vf}_{_{\! \barn(\veck)}}^{_{\, \veck}} \, \Big\vert \, \vf_{_{\! n(\veck)}}^{_{\, \veck}} \Big\rangle \quad ,
\eeq
where $\Phi_{\{ n \}} \in \mscrS_{_{\!\!\G}}$ and $\bar{\Phi}_{\{ \barn \}} \in \mscrS_{_{\!\!\barG}}\,$. A useful simplif\mbox{}ication comes from noting that, since $\sum \barn(\veck) < + \infty$ and $\sum n(\veck) < + \infty\,$, only a f\mbox{}inite number of terms in the numerical product on the right hand side are dif\mbox{}ferent from $\big\langle \bar{\vf}_{_{0}}^{_{\, \veck}} \mid \vf_{_{0}}^{_{\, \veck}} \big\rangle\,$. Moreover, a direct calculation using (\ref{varphi}) and (\ref{barvarphi}) shows that
\begin{equation} \label{Patty}
\Babs{\Big\langle \bar{\vf}_{_{\!0\phantom{(\veck)}}}^{_{\, \veck}} \!\!\! \Big\vert \, \vf_{_{\!0\phantom{(\veck)}}}^{_{\, \veck}} \!\!\!\! \Big\rangle} = \exp \bigg[ \! - \frac{1}{4} \, \Big( \big( \s_{_{\!\veck}} - \bars_{_{\!\veck}} \big)^{\!2} + \big( \t_{_{\!\veck}} - \bart_{_{\!\veck}} \big)^{\!2} \Big) \bigg] \quad ,
\end{equation}
which is dif\mbox{}ferent from zero for every $\veck \in \mcalR\,$. This implies that, indicating with $\mcalR^{\p} \subset \mcalR$ the set of wavevectors $\veck$ for which $\barn(\veck)$ and $n(\veck)$ do not both vanish, we can write
\begin{equation} \label{connection}
\Big\langle \bar{\Phi}_{\{ \barn \}} \, \Big\vert \, \Phi_{\{ n \}} \Big\rangle = \Bigg[ \, \prod_{\veck \in \mcalR^{\p}} \Big\langle \bar{\vf}_{_{\! \barn(\veck)}}^{_{\, \veck}} \, \Big\vert \, \vf_{_{\! n(\veck)}}^{_{\, \veck}} \Big\rangle \, \Big\langle \bar{\vf}_{_{\!0\phantom{(\veck)}}}^{_{\, \veck}} \!\!\! \Big\vert \, \vf_{_{\!0\phantom{(\veck)}}}^{_{\, \veck}} \!\!\!\! \Big\rangle^{\! -1}\, \Bigg] \, \Big\langle \bar{\Phi}_{\{ 0 \}} \, \Big\vert \, \Phi_{\{ 0 \}} \Big\rangle \quad .
\end{equation}
Since the cardinality of $\mcalR^{\p}$ is f\mbox{}inite, the term in square parenthesis is just a f\mbox{}inite factor. It follows that if the scalar product $\big\langle \bar{\Phi}_{\{ 0 \}} \, \big\vert \, \Phi_{\{ 0 \}} \big\rangle$ vanishes, then necessarily $\big\langle \bar{\Phi}_{\{ \barn \}} \, \big\vert \, \Phi_{\{ n \}} \big\rangle$ vanishes.

Note on this respect that from \eqref{Patty} we get
\begin{equation} \label{scalar product 0}
\Babs{\Big\langle \bar{\Phi}_{\{ 0 \}} \, \Big\vert \, \Phi_{\{ 0 \}} \Big\rangle} = \prod_{\veck \in \mcalR} \Babs{\Big\langle \bar{\vf}_{_{\!0\phantom{(\veck)}}}^{_{\, \veck}} \!\!\! \Big\vert \, \vf_{_{\!0\phantom{(\veck)}}}^{_{\, \veck}} \!\!\!\! \Big\rangle} = \exp \Bigg[ - \frac{1}{4} \sum_{\veck \in \mcalR} \Big( \big( \s_{_{\!\veck}} - \bars_{_{\!\veck}} \big)^{\!2} + \big( \t_{_{\!\veck}} - \bart_{_{\!\veck}} \big)^{\!2} \Big) \Bigg] \quad ,
\end{equation}
where
\begin{multline} \label{series}
\sum_{\veck \in \mcalR} \Big( \big( \s_{_{\!\veck}} - \bars_{_{\!\veck}} \big)^{\!2} + \big( \t_{_{\!\veck}} - \bart_{_{\!\veck}} \big)^{\!2} \Big) = \frac{1}{V} \sum_{\veck \in \mcalR} \frac{1}{\o_{k}^{3}} \,\, \sum_{i = 1}^{2} \sum_{j = 1}^{2} \bigg[ \, g_{_{i}} g_{_{j}} \cos \Big( \veck \, \cdot \, \big( \vecy_{i} \, -\, \vecy_{j} \big) \Big) + \\
+ \barg_{_{i}} \barg_{_{j}} \cos \Big( \veck \cdot \big( \vecz_{i} - \vecz_{j} \big) \Big) - 2 \, g_{_{i}} \barg_{_{j}} \cos \Big( \veck \cdot \big( \vecy_{i} - \vecz_{j} \big) \Big) \bigg] \quad .
\end{multline}
It is clear from (\ref{scalar product 0}) that $\big\langle \bar{\Phi}_{\{ 0 \}} \, \big\vert \, \Phi_{\{ 0 \}} \big\rangle$ can be dif\mbox{}ferent from zero only if the series (\ref{series}) converges. The crucial observation is that the series $\sum_{\veck \in \mcalR} \o_{k}^{-3}$ diverges, while $\sum_{\veck \in \mcalR} \o_{k}^{-3} \cos (\veck \cdot \vec{v})$ converges (although not absolutely) provided $\vec{v} \neq \vec{0}\,$. Since $\vecy_{_{1}} \neq \vecy_{_{2}}$ and $\vecz_{_{1}} \neq \vecz_{_{2}}\,$, the only terms in the right hand side of (\ref{series}) which may not converge are
\begin{equation} \label{square b}
\frac{1}{V} \sum_{\veck \in \mcalR} \frac{1}{\o_{k}^{3}} \,\, \Bigg[ \, \sum_{i = 1}^{2} \, g^{2}_{_{i}} + \sum_{i = 1}^{2} \, \barg^{2}_{_{i}} - 2 \, \sum_{i = 1}^{2} \sum_{j = 1}^{2} \, g_{_{i}} \barg_{_{j}} \cos \Big( \veck \cdot \big( \vecy_{i} - \vecz_{j} \big) \Big) \Bigg] \quad .
\end{equation}
For this sum to converge, the divergence coming from the $g_{_{i}}^{2}$ and $\barg_{_{i}}^{2}$ terms need to be compensated by an opposite divergence coming from the $g_{_{i}} \barg_{_{j}}$ terms, and for this to happen necessarily the argument of some of the cosines has to vanish. In fact, the argument of at most two cosines can vanish, and this happens when $\vecy_{_{1}} = \vecz_{_{1}}$ and $\vecy_{_{2}} = \vecz_{_{2}}$, or when $\vecy_{_{1}} = \vecz_{_{2}}$ and $\vecy_{_{2}} = \vecz_{_{1}}$. In the former case the series (\ref{square b}) reads
\begin{equation}
\frac{1}{V} \sum_{\veck \in \mcalR} \frac{1}{\o_{k}^{3}} \,\, \bigg[ \, (g_{_{1}} - \barg_{_{1}})^{2} + (g_{_{2}} - \barg_{_{2}})^{2} + \text{cosine terms} \, \bigg] \quad ,
\end{equation}
which converges if and only if $g_{_{1}} = \barg_{_{1}}$ and $g_{_{2}} = \barg_{_{2}}$, while in the latter case one gets
\begin{equation}
\frac{1}{V} \sum_{\veck \in \mcalR} \frac{1}{\o_{k}^{3}} \,\, \bigg[ \, (g_{_{1}} - \barg_{_{2}})^{2} + (g_{_{2}} - \barg_{_{1}})^{2} + \text{cosine terms} \, \bigg] \quad ,
\end{equation}
which converges if and only if $g_{_{1}} = \barg_{_{2}}$ and $g_{_{2}} = \barg_{_{1}}$. By ``cosine terms'' we mean terms proportional to $\cos (\veck \cdot \vec{v})$ for some $\vec{v} \neq \vec{0}\,$. Physically these two cases are the same, since the dif\mbox{}ference lies just in the way the sources of $\barG$ are labeled (that is, ``1'' and ``2'' instead of as ``2'' and ``1''). In both cases $\G$ and $\barG$ are equal as sets, that is $\G = \barG\,$. On the other hand, it is easy to see that we can never have convergence if the argument of only one cosine in (\ref{square b}) vanish. For example, if $\vecy_{_{1}} = \vecz_{_{1}}$ and $\vecy_{_{2}} \neq \vecz_{_{2}}$ the series reads
\begin{equation}
\frac{1}{V} \sum_{\veck \in \mcalR} \frac{1}{\o_{k}^{3}} \,\, \bigg[ \, (g_{_{1}} - \barg_{_{1}})^{2} + g_{_{2}}^{2} + \barg_{_{2}}^{2} + \text{cosine terms} \, \bigg] \quad ,
\end{equation}
which diverges (unless $g_{_{2}} = \barg_{_{2}} = 0$, in which case however there is only one source). In the cases $\{ \vecy_{_{1}} = \vecz_{_{2}}\,$, $\vecy_{_{2}} \neq \vecz_{_{1}} \}$, $\{ \vecy_{_{2}} = \vecz_{_{1}}\,$, $\vecy_{_{1}} \neq \vecz_{_{2}} \}$ and $\{ \vecy_{_{2}} = \vecz_{_{2}}\,$, $\vecy_{_{1}} \neq \vecz_{_{1}}\}$ the situation is analogous. We have therefore proved that the series (\ref{series}) converges if and only if $\G = \barG\,$, which by (\ref{scalar product 0}) means that $\big\langle \bar{\Phi}_{\{ 0 \}} \, \big\vert \, \Phi_{\{ 0 \}} \big\rangle$ vanishes whenever $\G \neq \barG\,$.

Recalling now that (\ref{connection}) holds whenever $\G \neq \barG\,$, we conclude that the scalar product $\big\langle \bar{\Phi}_{\{ \barn \}} \, \big\vert \, \Phi_{\{ n \}} \big\rangle$ vanishes whenever $\G \neq \barG\,$, so our claim is proved.

\subsection{Second orthogonality result}

Suppose $\G \neq \barG\,$. Similarly to the previous case, if $\sum n(\veck) < + \infty$ then only a f\mbox{}inite number of terms in the numerical product in (\ref{scalar product nm}) have $n(\veck) \neq 0\,$. Moreover, using (\ref{varphi}) and (\ref{barvarphi}) we get
\begin{equation} \label{scalar product n0}
\Babs{\Big\langle \bar{\vf}_{_{\barn\phantom{(}}}^{_{\, \veck}} \Big\vert \, \vf_{_{\!0\phantom{(\veck)}}}^{_{\, \veck}} \!\!\!\! \Big\rangle} = \exp \bigg[ \! - \frac{1}{4} \, \Big( \big( \s_{_{\!\veck}} - \bars_{_{\!\veck}} \big)^{\!2} + \big( \t_{_{\!\veck}} - \bart_{_{\!\veck}} \big)^{\!2} \Big) \bigg] \,\, \babs{\mscr{F} \big( \barn, \z_{_{\veck}} \big)} \quad ,
\end{equation}
where
\begin{equation} \label{zeta}
\z_{_{\veck}} = \frac{\bart_{_{\!\veck}} - \t_{_{\!\veck}} + i \, \big( \bars_{_{\!\veck}} - \s_{_{\!\veck}} \big)}{2}
\end{equation}
and
\begin{equation} \label{F function}
\mscr{F} \big( \barn \, , \z_{_{\veck}} \big) = \frac{1}{\sqrt{\pi}} \int_{- \infty}^{+ \infty} e^{- \big( \xi - \z_{_{\veck}} \big)^{2}} \, \frac{H_{\barn}(\xi)}{\sqrt{2^{\barn} \, \barn ! \,}} \,\, d\xi \quad .
\end{equation}
The result of the previous section is correctly recovered for $\barn = 0\,$. The integral in (\ref{F function}) can be evaluated explicitly \cite{Gradshteyn Ryzhik 7ed} and we get
\begin{equation}
\babs{\mscr{F} \big( \barn \, , \z_{_{\veck}} \big)} = \frac{1}{\sqrt{2^{\barn} \, \barn ! \,}} \,\, \Babs{ \, \bart_{_{\!\veck}} - \t_{_{\!\veck}} + i \, \big( \bars_{_{\!\veck}} - \s_{_{\!\veck}} \big)}^{\barn} \quad ,
\end{equation}
which implies that $\babs{\big\langle \bar{\vf}_{_{\barn\phantom{(}}}^{_{\, \veck}} \big\vert \vf_{_{0}}^{_{\, \veck}} \big\rangle}$ is dif\mbox{}ferent from zero. Calling $\mcalR^{\p\p} \subset \mcalR$ the set of wavevectors $\veck$ for which $n(\veck)$ does not vanish, we can therefore write
\begin{equation} \label{important}
\Big\langle \bar{\Phi}_{\{ \barn \}} \, \Big\vert \, \Phi_{\{ n \}} \Big\rangle = \Bigg[ \, \prod_{\veck \in \mcalR^{\p\p}} \Big\langle \bar{\vf}_{_{\! \barn(\veck)}}^{_{\, \veck}} \Big\vert \vf_{_{\! n(\veck)}}^{_{\, \veck}} \!\Big\rangle \, \Big\langle \! \bar{\vf}_{_{\! \barn(\veck)}}^{_{\, \veck}} \Big\vert \, \vf_{_{\!0\phantom{(\veck)}}}^{_{\, \veck}} \!\!\! \Big\rangle^{\! -1}\, \Bigg] \, \Big\langle \bar{\Phi}_{\{ \barn \}} \, \Big\vert \, \Phi_{\{ 0 \}} \Big\rangle \quad ,
\end{equation}
where the term in the square parenthesis is again a f\mbox{}inite factor.

If $\barn \geq 1\,$, it can be shown that
\begin{equation*}
\babs{\big\langle \bar{\vf}_{_{\barn\phantom{(}}}^{_{\, \veck}} \, \big\vert \vf_{_{\!0\phantom{(\veck)}}}^{_{\, \veck}} \!\!\! \big\rangle} \lessgtr \babs{\big\langle \bar{\vf}_{_{\!0\phantom{(\veck)}}}^{_{\, \veck}} \!\!\! \big\vert \vf_{_{\!0\phantom{(\veck)}}}^{_{\, \veck}} \!\!\! \big\rangle} \quad \text{when} \quad \big( \s_{_{\!\veck}} - \bars_{_{\!\veck}} \big)^{\!2} + \big( \t_{_{\!\veck}} - \bart_{_{\!\veck}} \big)^{\!2} \lessgtr 2 \sqrt[\barn]{\barn ! \, } \quad ,
\end{equation*}
and that $\babs{\big\langle \bar{\vf}_{_{\barn\phantom{(}}}^{_{\, \veck}} \, \big\vert \vf_{_{\!0\phantom{(\veck)}}}^{_{\, \veck}} \!\!\! \big\rangle}$ is always strictly smaller than one. In fact, introducing the family of functions 
\begin{equation}
h_{\barn}(x) = \frac{1}{\sqrt{2^{\barn} \, \barn ! \,}} \,\, e^{- \frac{x^{2}}{4}} \, x^{\barn} \quad ,
\end{equation}
it is easy to see that they achieve their maximum at $x = \sqrt{2 \, \barn \,}$ and that
\begin{equation}
h_{\barn}\big( \sqrt{2 \, \barn \,} \big) = \sqrt{\frac{\barn^{\barn}}{e^{\barn} \, \barn ! \,}} < \frac{1}{\sqrt[4]{2 \pi}} \quad ,
\end{equation}
where we used Stirling's formula \cite{Robbins 1955} to infer the inequality. This means that for $\barn \geq 0$ we have
\begin{equation} \label{scalar product n0 bound}
\Babs{\Big\langle \bar{\vf}_{_{\barn\phantom{(}}}^{_{\, \veck}} \Big\vert \, \vf_{_{\!0\phantom{(\veck)}}}^{_{\, \veck}} \!\!\!\! \Big\rangle} \leq \textup{max} \, \Bigg\{ \! \exp \bigg[ \! - \frac{1}{4} \, \Big( \big( \s_{_{\!\veck}} - \bars_{_{\!\veck}} \big)^{\!2} + \big( \t_{_{\!\veck}} - \bart_{_{\!\veck}} \big)^{\!2} \Big) \bigg] \, , \, \frac{1}{\sqrt[4]{2 \pi}} \Bigg\} \quad ,
\end{equation}
which implies
\beq
\label{scalar product n}
\Big\langle \bar{\Phi}_{\{ \barn \}} \, \Big\vert \, \Phi_{\{ 0 \}} \Big\rangle = \prod_{\veck \in \mcalR} \Big\langle \bar{\vf}_{_{\! \barn(\veck)}}^{_{\, \veck}} \Big\vert \, \vf_{_{0}}^{_{\, \veck}} \Big\rangle = 0 \quad .
\eeq
Since (\ref{important}) holds when $\G \neq \barG\,$, the relation (\ref{scalar product n}) implies that $\big\langle \bar{\Phi}_{\{ \barn \}} \, \big\vert \, \Phi_{\{ n \}} \big\rangle$ vanishes when $\G \neq \barG$ and $\sum n(\veck) < + \infty\,$, so our claim is proved.

\section{Factorization of the time evolution operator}
\label{App: Factorization}

Let us start from the expression (\ref{evolution operator}), to wit
\begin{equation} \label{evolution operator app}
U_{_{\!I}}^{\a}(t, - \infty) = \sum_{n = 0}^{+ \infty} \, \frac{1}{n!} \, \Big( \frac{g}{i} \Big)^{\! n} \! \int_{- \infty}^{t} d t_{_{1}} \ldots \int_{- \infty}^{t} d t_{_{n}} \, T \Big\{ \hH_{_{\!I}}^{\a}(t_{_{1}}) \cdots \hH_{_{\!I}}^{\a}(t_{_{n}}) \Big\} \quad ,
\end{equation}
and note that the number of dif\mbox{}ferent ways to perform $k$ contractions in a product of $n$ operators is
\beq
\sqcap_{\, k}^{\, n} = \frac{n!}{(2k)! \, (n - 2k)!} \, (2k - 1)!! = \frac{n!}{2^{k} \, k! \, (n - 2k)!} \quad ,
\eeq
where the double factorial $(2k - 1)!!$ expresses the number of dif\mbox{}ferent ways to completely contract a product of $2k$ operators. For the sake of clarity, let us split the sum in (\ref{evolution operator app}) into a sum over the even values of $n$ and a sum over the odd values of $n$
\begin{multline} \label{U even odd split}
U_{_{\!I}}^{\a}(t, - \infty) = \sum_{m = 0}^{+ \infty} \, \frac{1}{(2m)!} \, \Big( \frac{g}{i} \Big)^{\! 2m} \! \int_{- \infty}^{t} \!\! d t_{_{1}} \, \ldots \! \int_{- \infty}^{t} \!\! d t_{_{2m}} \, T \Big\{ \hH_{_{\!I}}^{\a}(t_{_{1}}) \cdots \hH_{_{\!I}}^{\a}(t_{_{2m}}) \Big\} + \\
+ \sum_{r = 0}^{+ \infty} \, \frac{1}{(2r + 1)!} \, \Big( \frac{g}{i} \Big)^{\! 2r + 1} \! \int_{- \infty}^{t} \!\! d t_{_{1}} \, \ldots \! \int_{- \infty}^{t} \!\! d t_{_{2r + 1}} \, T \Big\{ \hH_{_{\!I}}^{\a}(t_{_{1}}) \cdots \hH_{_{\!I}}^{\a}(t_{_{2r + 1}}) \Big\} \quad .
\end{multline}

Consider f\mbox{}irst the even part of the inf\mbox{}inite sum. When using the Wick theorem, the (double) integral of a contraction over the two time variables involved factors out from the $2m$-dimensional integral of the the time-ordered product. Moreover, the double integral gives the same result independently of which time variables (among $t_{_{1}} , \ldots , t_{_{2m}}$) are involved in the contraction. Therefore using the Wick theorem we get
\begin{multline} \label{Minerva}
\int_{- \infty}^{t} \!\! d t_{_{1}} \, \ldots \! \int_{- \infty}^{t} \!\! d t_{_{2m}} \, T \Big\{ \hH_{_{\!I}}^{\a}(t_{_{1}}) \cdots \hH_{_{\!I}}^{\a}(t_{_{2m}}) \Big\} = \sum_{k = 0}^{m} \, \sqcap_{\, k}^{\, 2m} \, \Bigg( \int_{- \infty}^{t} \!\! d \t_{_{1}} \!\! \int_{- \infty}^{t} \!\! d \t_{_{2}} \, \acontraction[1ex]{}{\hH_{_{\!I}}^{\a}}{(\t_{_{1}})}{\hH_{_{\!I}}^{\a}} \hH_{_{\!I}}^{\a}(\t_{_{1}}) \, \hH_{_{\!I}}^{\a}(\t_{_{2}}) \Bigg)^{\!\!k} \cdot \\
\cdot \, \Bigg( \int_{- \infty}^{t} \!\! d y_{_{1}} \, \ldots \! \int_{- \infty}^{t} \!\! d y_{_{2m - 2k}} \,\,  N \Big\{ \hH_{_{\!I}}^{\a}(y_{_{1}}) \cdots \hH_{_{\!I}}^{\a}(y_{_{2m - 2k}}) \Big\} \Bigg) \quad .
\end{multline}
Inserting this expression into the even part of (\ref{U even odd split}), we get the double sum
\begin{equation} \label{Heather}
\sum_{m = 0}^{+ \infty} \, \frac{1}{(2m)!} \, \Big( \frac{g}{i} \Big)^{\! 2m} \sum_{k = 0}^{m} \, \sqcap_{\, k}^{\, 2m} \, \Big( \mathrm{contraction} \Big)^{\! k} \Big( \text{normal ordered} \Big) \quad ,
\end{equation}
which can be seen as a sum over the set $\mcal{T} = \big\{ (k,m) \in \mbbN_{^{0}}^{2} : k \leq m \big\}$. Mapping this set onto $\mbbN_{^{0}}^{2}$ with the invertible map $(m,k) \to (m - k,k)$, and using the explicit expression for $\sqcap_{\, k}^{\, 2s + 2k}$, the sum (\ref{Heather}) can be written as
\begin{multline} \label{Heather 2}
\sum_{s = 0}^{+ \infty} \, \sum_{k = 0}^{+ \infty} \, \frac{1}{(2s)!} \, \Big( \frac{g}{i} \Big)^{\! 2s} \frac{1}{2^{k} \, k!} \, \Big( \frac{g}{i} \Big)^{\! 2k} \, \Bigg( \int_{- \infty}^{t} \!\! d \t_{_{1}} \!\! \int_{- \infty}^{t} \!\! d \t_{_{2}} \, \acontraction[1ex]{}{\hH_{_{\!I}}^{\a}}{(\t_{_{1}})}{\hH_{_{\!I}}^{\a}} \hH_{_{\!I}}^{\a}(\t_{_{1}}) \, \hH_{_{\!I}}^{\a}(\t_{_{2}}) \Bigg)^{\!\!k} \, \cdot \\
\cdot \, \Bigg( \int_{- \infty}^{t} \!\! d y_{_{1}} \, \ldots \! \int_{- \infty}^{t} \!\! d y_{_{2s}} \,\,  N \Big\{ \hH_{_{\!I}}^{\a}(y_{_{1}}) \cdots \hH_{_{\!I}}^{\a}(y_{_{2s}}) \Big\} \Bigg) \quad .
\end{multline}
It is apparent that the double sum factorizes into a sum over $s$ and one over $k$, and that the latter gives the exponential
\begin{equation*}
\sum_{k = 0}^{+ \infty} \, \frac{1}{k!} \, \Bigg( \frac{- g^{2}}{2} \! \int_{- \infty}^{t} \!\! d \t_{_{1}} \!\! \int_{- \infty}^{t} \!\! d \t_{_{2}} \, \acontraction[1ex]{}{\hH_{_{\!I}}^{\a}}{(\t_{_{1}})}{\hH_{_{\!I}}^{\a}} \hH_{_{\!I}}^{\a}(\t_{_{1}}) \, \hH_{_{\!I}}^{\a}(\t_{_{2}}) \Bigg)^{\!\!k}  = \exp \Bigg( \frac{- g^{2}}{2} \! \int_{- \infty}^{t} \!\! d \t_{_{1}} \!\! \int_{- \infty}^{t} \!\! d \t_{_{2}} \, \acontraction[1ex]{}{\hH_{_{\!I}}^{\a}}{(\t_{_{1}})}{\hH_{_{\!I}}^{\a}} \hH_{_{\!I}}^{\a}(\t_{_{1}}) \, \hH_{_{\!I}}^{\a}(\t_{_{2}}) \Bigg) \quad .
\end{equation*}

In the odd case an expression analogous to (\ref{Minerva}) holds, with $2m$ replaced by $2r + 1$ and the sum over $k$ extending from $0$ to $r\,$. The double sum $\sum_{r = 0}^{+ \infty} \, \sum_{j = 0}^{r}$ can be rearranged using the map $(r,k) \to (r - k,k)\,$, and again the ``contraction part'' factorizes into the same exponential as in the even case. These results taken together imply that $U_{_{\!I}}^{\a}(t, - \infty)$ factorizes into a ``contraction part'' and a ``normal ordered part'' according to 
\begin{multline} \label{evolution operator factorized app}
U_{_{\!I}}^{\a}(t, - \infty) = \exp \Bigg( \!\! - \frac{g^{2}}{2} \int_{- \infty}^{t} \!\! d \t_{_{1}} \!\! \int_{- \infty}^{t} \!\! d \t_{_{2}} \, \acontraction[1ex]{}{\hH_{_{\!I}}^{\a}}{(\t_{_{1}})}{\hH_{_{\!I}}^{\a}} \hH_{_{\!I}}^{\a}(\t_{_{1}}) \, \hH_{_{\!I}}^{\a}(\t_{_{2}}) \Bigg) \, \cdot \\
\cdot \Bigg( \sum_{n = 0}^{+ \infty} \, \frac{1}{n!} \, \Big( \frac{g}{i} \Big)^{\! n} \! \int_{- \infty}^{t} \!\! d t_{_{1}} \, \ldots \int_{- \infty}^{t} \!\! d t_{_{n}} \, N \Big\{ \hH_{_{\!I}}^{\a}(t_{_{1}}) \cdots \hH_{_{\!I}}^{\a}(t_{_{n}}) \Big\} \Bigg) \quad .
\end{multline}

\end{document}